\documentstyle[11pt]{article}     

\expandafter\ifx\csname amssym.def\endcsname\relax \else\endinput\fi
\expandafter\edef\csname amssym.def\endcsname{%
       \catcode`\noexpand\@=\the\catcode`\@\space}
\catcode`\@=11
\def\undefine#1{\let#1\undefined}
\def\newsymbol#1#2#3#4#5{\let\next@\relax
 \ifnum#2=\@ne\let\next@\msafam@\else
 \ifnum#2=\tw@\let\next@\msbfam@\fi\fi
 \mathchardef#1="#3\next@#4#5}
\def\mathhexbox@#1#2#3{\relax
 \ifmmode\mathpalette{}{\m@th\mathchar"#1#2#3}%
 \else\leavevmode\hbox{$\m@th\mathchar"#1#2#3$}\fi}
\def\hexnumber@#1{\ifcase#1 0\or 1\or 2\or 3\or 4\or 5\or 6\or 7\or 8\or
 9\or A\or B\or C\or D\or E\or F\fi}

\font\tenmsa=msam10
\font\sevenmsa=msam7
\font\fivemsa=msam5
\newfam\msafam
\textfont\msafam=\tenmsa
\scriptfont\msafam=\sevenmsa
\scriptscriptfont\msafam=\fivemsa
\edef\msafam@{\hexnumber@\msafam}
\mathchardef\dabar@"0\msafam@39
\def\dashrightarrow{\mathrel{\dabar@\dabar@\mathchar"0\msafam@4B}}
\def\dashleftarrow{\mathrel{\mathchar"0\msafam@4C\dabar@\dabar@}}

\def\ulcorner{\delimiter"4\msafam@70\msafam@70 }
\def\urcorner{\delimiter"5\msafam@71\msafam@71 }
\def\llcorner{\delimiter"4\msafam@78\msafam@78 }
\def\lrcorner{\delimiter"5\msafam@79\msafam@79 }
\def\yen{{\mathhexbox@\msafam@55}}
\def\checkmark{{\mathhexbox@\msafam@58}}
\def\circledR{{\mathhexbox@\msafam@72}}
\def\maltese{{\mathhexbox@\msafam@7A}}
\def\circledS{{\mathhexbox@\msafam@73}}

\font\tenmsb=msbm10
\font\sevenmsb=msbm7
\font\fivemsb=msbm5
\newfam\msbfam
\textfont\msbfam=\tenmsb
\scriptfont\msbfam=\sevenmsb
\scriptscriptfont\msbfam=\fivemsb
\edef\msbfam@{\hexnumber@\msbfam}
\def\Bbb#1{{\fam\msbfam\relax#1}}
\def\widehat#1{\setbox\z@\hbox{$\m@th#1$}%
 \ifdim\wd\z@>\tw@ em\mathaccent"0\msbfam@5B{#1}%
 \else\mathaccent"0362{#1}\fi}
\def\widetilde#1{\setbox\z@\hbox{$\m@th#1$}%
 \ifdim\wd\z@>\tw@ em\mathaccent"0\msbfam@5D{#1}%
 \else\mathaccent"0365{#1}\fi}
\font\teneufm=eufm10
\font\seveneufm=eufm7
\font\fiveeufm=eufm5
\newfam\eufmfam
\textfont\eufmfam=\teneufm
\scriptfont\eufmfam=\seveneufm
\scriptscriptfont\eufmfam=\fiveeufm
\def\frak#1{{\fam\eufmfam\relax#1}}

\csname amssym.def\endcsname

\makeatletter
\def\section{\@startsection {section}{1}{\z@}{-3.5ex plus -1ex minus 
 -.2ex}{2.3ex plus .2ex}{\large\sc}}
\def\subsection{\@startsection{subsection}{2}{\z@}{-3.25ex plus -1ex minus 
 -.2ex}{1.5ex plus .2ex}{\normalsize\sc}}
\makeatother

\makeatletter
\@addtoreset{equation}{section}                        

\makeatother

\newcommand{\nc}{\newcommand}
\newcommand{\rnc}{\renewcommand}

\nc{\be}{\begin{equation}}
\nc{\ee}{\end{equation}}
\nc{\bea}{\begin{eqnarray}}
\nc{\eea}{\end{eqnarray}}

\nc{\trac}[2]{{\textstyle\frac{#1}{#2}}}

\nc{\ex}[1]{\mbox{e}^{\,\textstyle#1}}

\nc{\CC}{\Bbb{C}}
\nc{\PP}{\Bbb{P}}
\nc{\RR}{\Bbb{R}}
\nc{\ZZ}{\Bbb{Z}}
\nc{\II}{\Bbb{I}}
\nc{\EE}{\Bbb{E}}
\nc{\SS}{\Bbb{S}}

\rnc{\a}{\alpha}
\nc{\al}{\a^{l}}
\rnc{\d}{\delta}
\nc{\ga}{\gamma}
\nc{\la}{\lambda}
\nc{\lal}{\la_{l}}
\nc{\f}{\phi}
\nc{\fb}{\bar{\phi}}
\nc{\p}{\psi}
\def\pb{\bar{\p}}
\nc{\e}{\eta}
\nc{\eb}{\bar{\eta}}
\rnc{\c}{\chi}
\nc{\eps}{\epsilon}
\rnc{\t}{\theta}
\nc{\tb}{\bar{\theta}}
\nc{\om}{\omega}

\rnc{\P}{\Psi}
\nc{\pl}{\P_{L}}
\nc{\pdr}{\P^{\dag}_{R}}
\nc{\G}{\Gamma}

\nc{\sig}{\sigma}
\nc{\sk}{\sigma_{k}}
\nc{\sa}{\sigma_{a}}
\nc{\Bb}{\bar{B}}

\nc{\symx}{\circledS}

\nc{\Q}{\bar{Q}}
\nc{\M}{{\cal M}}                          
\nc{\C}{{\cal A}/{\cal G}}                
\nc{\A}[1]{{\cal A}^{#1}/{\cal G}^{#1}}  
\nc{\RC}{{\cal R}_{\C}}                 
\nc{\RM}{{\cal R}_{\M}}                
\nc{\RX}{{\cal R}_{X}}
\nc{\RY}{{\cal R}_{Y}}

\nc{\ad}{\mathop{\mbox{ad}}\nolimits}
\nc{\tr}{\mathop{\mbox{tr}}\nolimits}
\nc{\Tr}{\mathop{\mbox{Tr}}\nolimits}
\nc{\Det}{\mathop{\mbox{Det}}\nolimits}
\nc{\rk}{\mathop{\mbox{rk}}\nolimits}
\nc{\diag}{\mbox{diag}}
\nc{\ra}{\rightarrow}
\nc{\Ra}{\Rightarrow}
\nc{\LRa}{\Leftrightarrow}
\nc{\ot}{\otimes}
\rnc{\ss}{\subset}
\nc{\nul}{\noindent\underline}
\nc{\non}{\nonumber\\}
\rnc{\S}{\Sigma}
\nc{\tp}{2\pi i}
\nc{\del}{\partial}
\nc{\dbar}{\bar{\del}}
\nc{\dx}{\dot{x}}
\rnc{\lg}{\frak{g}}

\nc{\mat}[4]{\left(\begin{array}{cc}#1&#2\\#3&#4\end{array}\right)}
 
\nc{\subs}[1]{{\vspace*{0.5cm}}%
{\noindent\underline{#1}}{\addcontentsline{toc}{subsection}{\sc #1}}%
{\vspace*{0.3cm}}}

\begin{document}
\global\parskip=4pt

\makeatletter
\begin{titlepage}
\rightline{{\small E}N{\large S}{\Large L}{\large A}P{\small P}-L-630/96}
\rightline{IC/96/262}
\rightline{PAR-LPTHE 96/53}
\rightline{hep-th/9612143}
\begin{center}
{\LARGE\sc Aspects~of~$N_{T}\geq 2$~Topological\\[4mm]
Gauge Theories and D-Branes}
\vskip .3in
\begin{tabular}{cc}
{\sc Matthias Blau}\footnotemark
&
{\sc George Thompson}\footnotemark 
\\[.1in]
LPTHE-{\sc enslapp}\footnotemark
&
ICTP\\
ENS-Lyon,
46 All\'ee d'Italie
&
P.O. Box 586\\
F-69364 Lyon CEDEX 07, France
&
34014 Trieste, Italy\\
\end{tabular}
\end{center}
\addtocounter{footnote}{-2}%
\footnotetext{e-mail: mblau@enslapp.ens-lyon.fr; supported 
by EC Human Capital and Mobility Grant ERB-CHB-GCT-93-0252}
\addtocounter{footnote}{1}%
\footnotetext{e-mail: thompson@ictp.trieste.it}
\addtocounter{footnote}{1}%
\footnotetext{URA 14-36 du CNRS, associ\'ee \`a l'E.N.S. de Lyon,
et \`a l'Universit\'e de Savoie}
\vskip .10in
\begin{small}
\noindent 
We comment on various aspects of topological gauge theories possessing
$N_{T}\!\geq\!2$ topological symmetry: 
\begin{enumerate}
\item We show that the construction of Vafa-Witten and Dijkgraaf-Moore of
`balanced' topological field theories is equivalent to an earlier
construction in terms of $N_{T}\!=\!2$ superfields inspired by
supersymmetric quantum mechanics.

\item We explain the relation between topological field theories
calculating signed and unsigned sums of Euler numbers of moduli
spaces.

\item We show that the topological twist of $N\!=\!4$ $d\!=\!4$
Yang-Mills theory recently constructed by Marcus is formally a
deformation of four-dimensional super-BF theory.

\item We construct a novel $N_{T}\!=\!2$ topological twist of $N=4$ $d=3$
Yang-Mills theory, a `mirror' of the Casson invariant model, with 
certain unusual features (e.g.\ no bosonic scalar field and hence no 
underlying equivariant cohomology) 

\item We give a complete classification of the topological twists of
$N\!=\!8$ $d\!=\!3$ Yang-Mills theory and show that they are realised
as world-volume theories of Dirichlet two-brane instantons wrapping
supersymmetric three-cycles of Calabi-Yau three-folds and
$G_{2}$-holonomy Joyce manifolds.

\item We describe the topological gauge theories associated to D-string
instantons on holomorphic curves in K3s and Calabi-Yau 3-folds.
\end{enumerate}
\end{small}
 
\end{titlepage}
\makeatother

\begin{small}
\tableofcontents
\end{small}
\setcounter{footnote}{0}
\section{Introduction}

Recently, topological field theories with extended $N_{T}>1$ topological
symmetries have appeared in various contexts, e.g.\ in the discussion
of S-duality in supersymmetric gauge theories \cite{vw,bjsv}, as world 
volume theories of Dirichlet $p$-branes in string theory \cite{bsv}, and
in a general discussion of `balanced' or critical topological theories
\cite{dm}. Here we will comment on, explain, or expand on various aspects
of these theories, thus complementing the already existing discussions
of such models in the literature. 

Let us begin by recalling that, in contrast to cohomological topological
field theories with an extended (in particular $N_{T}>2$) topological
symmetry, $N_{T}\!=\!1$ topological
theories (the most prominent example being Donaldson-Witten or topological
Yang-Mills theory \cite{ewdon}) are reasonably well understood (see e.g.\
\cite{bbrt} for a review). They typically capture 
the deformation complex of some underlying moduli problem, their partition
function is generically zero because of fermionic zero modes while
correlation functions correspond to intersection pairings on the 
moduli space. 

In \cite{btmq,btsqm} we investigated in some detail topological gauge
theories of a particular kind possessing an $N_{T}\!=\!2$ topological
symmetry. These theories typically capture the de Rham complex and
Riemannian geometry of some underlying moduli space, and they have the
characteristic property of being `critical', i.e.\ of possessing a
generically non-zero partition function equalling the Euler number of
the moduli space.  These properties are obviously reminiscent of
supersymmetric quantum mechanics and in \cite{btmq,btsqm} we made this
connection precise using the Mathai-Quillen formalism \cite{mq,aj,mbmq}
and an $N_{T}\!=\!2$ superfield version of a new variant of supersymmetric
quantum mechanics based on the Gauss-Codazzi equations.

More recently there has been renewed interest in topological field theories
possessing extended $N_{T}\!=\!2$ topological symmetries. For example, in
\cite{vw} Vafa and Witten considered a particular topologically twisted
version of $N\!=\!4$ $d\!=\!4$ super-Yang-Mills theory to perform a strong
coupling test of S-duality of the underlying supersymmetric theory. By
analogy with the nomenclature for supersymmetric sigma models \cite{ewab}, 
we shall (as suggested in \cite{nm}) refer to this model as the A-twist 
of $N=4$ $d=4$ Yang-Mills theory or simply as the A-model. 
The partition function of this topological gauge theory equals the Euler
number of instanton moduli space or, more precisely, in the case that
the instanton moduli space is not connected, the sum of the seperate
Euler numbers disregarding the relative orientations.

Along similar lines, in \cite{cmr} a topological
string theory calculating the Euler number of the
(Hurwitz) moduli space of branched covers was constructed and shown to
reproduce the large $N$ expansion of two-dimensional Yang-Mills theory 
obtained by Gross and Taylor \cite{gt}.

The constructions of \cite{vw} and \cite{cmr} have recently been 
generalized by Dijkgraaf and Moore \cite{dm} who christened
these theories `balanced' topological field theories and  analyzed
in detail the underlying $N_{T}\!=\!2$ equivariant cohomology. 

On the face of it the constructions of $N_{T}\!=\!2$ theories given in
\cite{dm} and \cite{btmq,btsqm} respectively appear to be quite
different. One purpose of this paper is to show that they are actually
completely equivalent. This observation sheds light on both approaches
as on the one hand it provides a cohomological underpinning to the
construction of \cite{btmq,btsqm} while on the other hand it gives an
{\em a priori} explanation in terms of Riemannian geometry for why
balanced topological field theories calculate Euler numbers.

We also pause to explain the relation between theories calculating
Euler numbers with and without relative signs, as this turns out
to be particularly transparent from the supersymmetric quantum mechanics
point of view and is something we should have stated more clearly 
in \cite{btmq}. 

A somewhat different kind of $N_{T}\!=\!2$ symmetry, not falling into the
above scheme of things, appears 
in `the other' topological twist of $N\!=\!4$ Yang-Mills theory (the
B-model), mentioned
in \cite{yamron,vw} and  recently constructed explicitly 
by Marcus \cite{nm}. We will show that the B-model is formally
a deformation of four-dimensional super-BF theory \cite{bbt,bbrt},  
i.e.\ the four-dimensional 
cohomological field theory (formally) describing moduli spaces of
real flat connections.\footnote{A slightly different description of
this theory has been obtained by Labastida and Lozano (J.M.F.\ 
Labastida, private
communication).} The deformation in question deforms the
cotangent bundle of the moduli space of real flat connections to the
moduli space of complex flat connections and one expects that this
does not change the topological aspects of the theory. One can then
understand the $N_{T}\!=\!2$ topological symmetry as a consequence of the
complex nature of the moduli space at this particular point in the
deformation space of super-BF theory.

This is reminiscent of the extended topological symmetries that can
arise when $N_{T}\!=\!1$ theories are formulated on manifolds with reduced
holonomy groups, as in Donaldson-Witten theory on a K\"ahler manifold.
In both cases this leads to a (Dolbeault) refinement of the original
(de Rham) intersection theory model but does not change 
the theory in other respects.

Continuing on our stroll through the zoo of $N_{T}\!=\!2$ theories, we
construct a novel topological gauge theory in three dimensions by 
twisting the $N\!=\!4$ $d\!=\!3$ theory, the dimensional reduction of
$N\!=\!1$ $d\!=\!6$ or $N\!=\!2$ $d\!=\!4$ theory, 
by the internal Lorentz group
$SU(2)$. As a consequence, the potential three scalars (the 
internal components of the gauge field) transform as a vector $V$ 
and this model has no bosonic scalars in its spectrum. Hence,
as there is in particular no scalar ghost for ghost, the topological
symmetry of this model cannot possibly correspond to equivariant 
cohomology and indeed we find that here the topological symmetries are 
strictly (and not only equivariantly, as in all the models above)
nilpotent and anti-commuting. Another novel feature of this model
is that while it is a cohomological gauge theory it permits
bosonic Wilson loops (of $A-iV$) as fully topological observables. 

Along rather different lines,  
Bershadsky, Sadov and Vafa \cite{bsv} have recently made the beautiful
observation that topologically
twisted gauge theories appear completely naturally as (low-energy effective)
world-volume
theories of Dirichlet $p$-brane instantons in string theory (for
recent reviews of these matters see \cite{jptasi,jsictp}). In particular,
they showed that the three (partial) topological twists of $N=4$ $d=4$
Yang-Mills theory \cite{yamron,vw,nm} are realized in this way.

Motivated by this, we classify the topological and partial topological twists
of $N\!=\!8$ $d\!=\!3$ (the dimensional reduction of $N\!=\!4$ $d\!=\!4$) 
Yang-Mills theory. Once again one finds that all of them are naturally
realized as worldvolume theories of Dirichlet two-brane instantons, on
special Lagrangian submanifolds of Calabi-Yau three-folds and associative
submanifolds of $G_{2}$-holonomy seven-manifolds respectively. 
In the same spirit, we also describe the topological gauge theories 
associated to D-string
instantons wrapping holomorphic curves in K3s and Calabi-Yau three-folds
respectively. 

We also expand slightly on the discussion given in \cite{bsv}
by showing that in all these theories the rotation group of the
uncompactified dimensions (in the string interpretation) is realized
as a global symmetry group of the world-volume theories - as required by 
the geometrical interpretation. Conversely, demanding such a symmetry
determines the relevant class of branchings of the R-symmetry group
and thus provides a short-cut to the construction of the 
desired topological theory via twisting. 

Finally, we want to point out that to a certain extent similar constructions
are possible in $d=5$ and $d=6$. For example, there should be a close
relation between twisted theories on $M_{4}\times T^{2}$, the 
$(4+2)$-dimensional
generalization of the $(4+1)$-dimensional quantum mechanics model of 
\cite{btmq,btsqm} and section 2, and the considerations in \cite{dvv}.
It is also straightforward to construct a topological gauge theory in
$d=5$. It appears to be related to the constructions and considerations
in \cite{lmns}. These issues are currently under investigation.

This paper is organized as follows. In section 2 we recall the 
supersymmetric quantum mechanics based construction of topological
field theories \cite{btmq,btsqm,mbmq} and establish the link with
the approaches of \cite{vw,cmr,dm}. In section 3, we present a detailed
comparison of the topological superfield approach of \cite{btmq} with
the approach of Dijkgraaf and Moore \cite{dm} based on $N_{T}\geq 1$
extended equivariant cohomology, establishing their equivalence.
In section 4 we discuss a simplified quantization of $d=4$ super-BF
theory (section 4.1), show that it can be deformed to the action given in
\cite{nm} (section 4.2), and introduce a novel topological twist
of $N=4$ $d=3$ Yang-Mills theory. In section 5, we classify the 
topological twists of $N=8$ $d=3$ Yang-Mills theory (section 5.1),
show that they are realized as world-volume theories of wrapped
two-brane instantons on supersymmetric three-cycles (section 5.2),
and extend the discussion to topological gauge theories associated with
D-string instantons wrapping holomorphic curves (section 5.3).

\section{$N_{T}\!=\!2$ Topological Field Theories and Euler Numbers} 

It is well known that supersymmetric quantum mechanics \cite{sqm} (SQM)
can be regarded as the archetype of a cohomological field theory (for a
review of SQM from this point of view see \cite{bbrt}). In particular,
what is commonly known as the $N\!=\!1/2$ model, calculating the index of
the Dirac operator, generalizes to cohomological field theory with an
$N_{T}\!=\!1$ topological supersymmetry\footnote{In topological field
theories it is more natural to count real scalar supercharges instead
of spinors} describing the geometry of the Atiyah-Singer universal
bundle and intersection theory on moduli spaces. 

The partition function
of the $N\!=\!1$ SQM model, on the other hand, equals the Euler number of
the target space, and it is thus natural to suspect that this
generalizes to $N_{T}\!=\!2$ topological field theories calculating the
Euler number of some moduli space. How this can be accomplished was explained
in detail in \cite{btmq,btsqm} where the structure of topological gauge 
theories with the desired properties was examined from several different
points of view. In particular, it was shown that actions for $N_{T}\!=\!2$
superfields encode the Riemannian geometry of the field space, leading 
via localization and the Gauss-Bonnet theorem to the Euler number of a
prescribed moduli space. We will briefly review this construction below.

Recently, an alternative approach to this kind of topological field theories,
based on $N_{T}\!=\!2$ equivariant cohomology,  was presented in \cite{dm}, 
generalizing the procedure employed in \cite{vw,cmr}. By comparing the
field content and action of the resulting theories with those obtained in
\cite{btmq,btsqm} (and their - straightforward - adaptation to sigma models),
it is readily seen that the two constructions are equivalent. 
However, it
is instructive to also explain this in terms of the relation between the
(seemingly rather different) superfields appearing in these two approaches,
and we shall do so in detail in section 3.

One of the properties the SQM-inspired approach \cite{btmq,btsqm} makes
particularly transparent is the relation between theories counting
Euler numbers (or solutions of equations) with or without signs (recall
that constructing topological field theories accomplishing the latter
was the main motivation for the `cofield' construction in
\cite{vw,cmr,dm}). This will be explained in section 2.4.        

\subsection{Review: Variants of Supersymmetric Quantum Mechanics}

The principal aim of \cite{btmq,btsqm} was to explain a construction of
topological gauge theories formally calculating the Euler character of 
a specified moduli space $\M$ of connections. In its most pragmatic 
incarnation,
this construction is expressed in terms of $N_{T}\!=\!2$ superfields and an
action $S_{\M}$ which consists of a kinetic term and a supersymmetric delta
function constraint onto the desired moduli space. The relation between
this approach and that by Vafa and Witten \cite{vw} and Dijkgraaf and Moore
\cite{dm} based on equivariant cohomology will be explained in the next 
section. 

To set the stage for this, however, in the following we will
briefly recall the considerations based on supersymmetric quantum mechanics
\cite{btsqm} which inspired the superfield formulation
as they provide an {\em a priori\/} explanation for why the topological gauge
theories constructed in this manner do what they are supposed to do. 
These remarks should also make it clear that the construction can be
straightforwardly adapted to e.g.\ topological sigma models or topological
gravity.

Let us start by recalling that the partition function of 
$N\!=\!1$ ($N_{T}\!=\!2$)
supersymmetric quantum mechanics with target space a Riemannian manifold
$(X,g)$, whose action is schematically of the form
\be
S_{X} = \int\! dt\; g_{\mu\nu}\dx^{\mu}\dx^{\nu} + \mbox{superpartners}\;\;,
\label{2}
\ee
equals the Euler nuber of $X$,
\be
Z(S_{X}) = \c(X)\;\;.
\ee
This can be seen in several ways \cite{sqm}, for instance by using the 
relation between the Witten index and the index of the de Rham complex
of $X$.
The partition function $Z(S_{X})$ can also be evaluated explicitly to give
a path integral proof of the Gauss-Bonnet theorem which expresses $\c(X)$
as an integral over $X$ of the Pfaffian of the curvature form $\RX$ of $X$,
\be
Z(S_{X})=\c(X)=\int_{X}\mbox{Pf}(\RX)\label{5}\;\;.
\ee
The crucial fact responsible for the reduction of the integral over the
loop space $LX$ of $X$ (the path integral) to an integral over $X$ (the
Gauss-Bonnet integral) is the (topological) supersymmetry which ensures
that only the Fourier zero modes (e.g.~$\dx=0$) of the fields are
relevant, the contributions from the other modes cancelling identically
between the bosonic and fermionic fields.

It is the analogue in infinite dimensions of this observation that permits 
one to construct topological gauge theories in $d$ (instead of $(d+1)$) 
dimensions from supersymmetric quantum mechanics on the space $\A{d}$ of 
gauge orbits of connections in $d$ dimensions. 

As they stand, the partition function and the right hand side of
(\ref{5}) do not make sense for infinite dimensional target spaces. 
There are, however, two refinements of the action (\ref{2})
which turn out to have meaningful counterparts on $\C$ \cite{btsqm}. 
The first of these
involves a choice of potential $V(x)$ on $X$, the corresponding action
being
\be
S_{X,V} =  \int\! dt\; g_{\mu\nu}
(\dx^{\mu}+g^{\mu\rho}\del_{\rho}V(x))
(\dx^{\nu}+g^{\nu\sigma}\del_{\sigma}V(x))
+ \mbox{superpartners}\;\;.
\label{6}
\ee
The partition function $Z(S_{X,V})$ localizes to an integral over the
set of critical points of $V$. There is also an intermediate localization
\cite{btmq,mbmq} which reproduces the finite-dimensional Mathai-Quillen
formalism \cite{mq} (and hence that of \cite{aj,vw,dm} in the field
theoretic setting).

In the case that the critical points of $V$ are isolated
and non-degenerate one arrives at the classical Poincar\'e-Hopf-Morse
theorem 
\be
\c(X)=\sum_{x_{k}: dV(x_{k})=0}(\pm 1)\label{7}\;\;.
\ee
If the critical points are not isolated then, by a combination of the 
arguments leading to (\ref{5}) and (\ref{7}), one finds
\be
\c(X)=\sum_{(k)}\c(X_{V}^{(k)})\label{8}\;\;,
\ee
where the $X_{V}^{(k)}$ are the connected components of the critical point
set of $V$ and once again the sum is to be taken with signs depending on
the relative orientations of the critical submanifolds induced by the
gradient flow. 

The relevance of this for our purposes is that the right hand side
of (\ref{8}) may be well defined, even if $X$ is infinite dimensional, 
provided that $X_{V}$ is finite dimensional. In that case $\c(X_{V})$ is
well defined and can be regarded as a regularized Euler number of $X$
(this is the point of view adopted in \cite{aj}). 
The advantage of the present construction 
is that it permits an {\em a priori} identification of
this $V$-dependent regularized Euler number of $X$ with the Euler number 
of $X_{V}$ \cite{btmq,mbmq}. 

Although this looks like a satisfactory state of affairs, we may not always 
be so fortunate to have a potential at our disposal whose critical points
define precisely the (moduli) subspace $Y\subset X$ we are interested in. 
In fact, it follows from (\ref{8}) that in finite dimensions 
$\c(Y)=\c(X)$ is a necessary condition for this to be possible. 

To motivate the following construction, recall that the
classical Gauss-Co\-daz\-zi equations 
express the intrinsic curvature $\RY$ of $Y$ (with the induced metric)
in terms of $\RX$ restricted to $Y$ and a term quadratic in the extrinsic 
curvature $K_{Y}$ of $Y$.  
The construction of an action $S_{Y\ss X}$ calculating $\c(Y)$  
will be modelled on this decomposition of $\RY$, i.e.\
it will consist of the action $S_{X}$
(\ref{2}) plus a supersymmetric 
Lagrange multiplier term enforcing the restriction to
$Y\ss X$. 
Concretely, one introduces an $N_{T}\!=\!2$ coordinate 
superfield $X(t,\t^{m})=x(t) + \ldots$
(where $\t^{m}=(\t,\tb)$ are Grassmann odd scalars) and,
assuming that $Y$ is (locally) defined by $F^{a}(x)=0$,
$N_{T}\!=\!2$ Lagrange multiplier fields
$\Lambda_{a}(t,\t^{m}) =  \lambda_{a}(t) + \ldots$ 
and chooses the action to be
\be
S_{Y\ss X}=S_{X}+\a\int dt\int d\t\,d\tb 
\Lambda_{a}(t,\t,\tb)F^{a}(X(t,\t,\tb))\label{13}\;\;,
\ee
so that the integration over the $\Lambda_{a}$ imposes the superconstraints 
$F^{a}(X)=0$. 
The argument leading to the elimination of the 
non-constant modes is not affected by the addition of this term and 
provided that there are no $\lambda$ zero modes the 
evaluation of the partition function and the Gauss-Bonnet theorem 
applied to $\RY$ lead to
\be
Z(S_{Y\ss X})=\c(Y)\label{16}\;\;,
\ee
now valid for arbitrary submanifolds $Y\ss X$ (not necessarily of the form
$X_{V}$). As usual in topological field theories, the construction works 
equally well when the delta function
constraint is replaced by a Gaussian, i.e.\ when one adds a term
$\sim \Lambda^{2}$ to the action. 
This is the generalization required to be able to apply 
supersymmetric quantum mechanics to spaces of connections in some sort
of generality. In fact, as has been shown in \cite{btmq,btsqm} and we will
recall in the following, the extension to infinite dimensional target 
spaces like
spaces of maps or metrics or connections is now fairly immediate. 

\subsection{Actions with Potentials and Action Potentials}

Let us, for concreteness, consider the case of gauge theories where
$X={\cal A}$ or $X=\C$, the space of connections on some manifold $M$.
A metric on $M$ induces a natural metric on ${\cal A}$ and then a metric
$g_{\C}$ on $\C$ via the horizontal projectors $h_{A}=\mbox{id} - d_{A}
(d_{A}^{*}d_{A})^{-1}d_{A}^{*}$. As a first step, in analogy with 
(\ref{2}), one might like to consider the action
\be
S_{\C} = \int\!dt\;g_{\C}(\dot{A},\dot{A}) + \mbox{superpartners}\;\;.
\ee
However, this does not yet give a well-defined action. But let us now, in
turn, consider the two refimements of the quantum mechanics action 
discussed above. In order to illustrate the field theoretic version
of the action $S_{X,V}$ (\ref{6}), we consider the space of gauge orbits
$\A{3}$ in $d\!=\!3$ dimensions. Then a natural candidate for $V$ is the
Chern-Simons functional $V_{CS}(A)$ (that this is not quite a well-defined
function on $\A{3}$, because of large gauge transformations, is not a problem
as only its derivative enters in the construction of the action which is a
well-defined closed but not exact one-form on $\A{3}$). Then the counterpart
of (\ref{6}) is
\be
S_{\A{3},V_{CS}} = \int\!dt\;g_{\A{3}}(\dot{A}-*F_{A},\dot{A}-*F_{A})
+ \mbox{superpartners}\;\;.
\ee
A priori, this looks like a $d\!=\!(3+1)$ dimensional field theory, and indeed 
it can be
shown \cite{btsqm} that this is precisely the action of Donaldson-Witten
theory on a four-manifold of the form $M\times S^{1}$. However, we already
know that all the time-dependent modes can be eliminated from this action,
leaving us with a three-dimensional field theory of the form
\be
S_{\A{3},V_{CS}}\ra \int_{M} F_{A}*F_{A} + \mbox{superpartners}
\label{qmsbf3}
\ee
calculating the Euler number
of the moduli space $\M^{(3)}$ of flat connections on $M$, the critical
points of $V_{CS}$. 

Of course, this procedure of constructing a field theory
in $d$ dimensions from another one in $d+1$ is somewhat indirect and the
superfield formalism introduced in \cite{btmq} is a way of bypassing the
auxiliary $(d+1)$-dimensional field theory and constructing directly
the resulting $d$-dimensional action. In the $N_{T}\!=\!2$ superfield language,
the action then consists of a universal kinetic term for 
$N_{T}\!=\!2$ theories, corresponding to $S_{\C}$ (in terms of the $N_{T}=2$ 
superfields to be discussed in
section 3, this is simply the term $\int d\t^{1}d\t^{2}{\cal F}_{1}*{\cal
F}_{2}$, where ${\cal F}_{m}$ is given in (\ref{calfm})), and an 
$N_{T}\!=\!2$ Chern-Simons action, so that,
in the parlance of \cite{dm}, $V_{CS}$ (or whatever potential one chooses)
is the bosonic part of the action potential.  

\subsection{Supersymmetric Quantum Mechnics and the Cofield Construction}

To illustrate the Gauss-Codazzi version (\ref{13}) of supersymmetric quantum
mechanics and its uses, let us try to construct a theory of flat connections
in $d\!=\!2$. In this case, there is no potential
constructed from the gauge fields alone whose critical points would make up
the desired moduli space. However, we can extend the space of fields and
consider `potentials' of the form $\int BF_{A}$ where $B$ is an adjoint
scalar field. The $N_{T}\!=\!2$ supersymmetric extension of
this is precisely the Lagrange multiplier construction of (\ref{13}), and
the resulting theory is described by the action 
\be
S_{\M\ss\C} = S_{\C} + \int\!dt\;BF_{A} + \mbox{superpartners}
\label{d24}
\ee
This reduces firstly to the constant modes and thus to the superfield
construction of \cite{btmq} in which the first term of (\ref{d24}) 
is replaced by the universal kinetic term for $N_{T}\!=\!2$ theories and
the time-dependence of the second term is dropped.
It subsequently reduces to the desired moduli space 
$\M$ provided that  there are no $B$
moduli, i.e.\ provided that one has a `vanishing theorem' \cite{vw}. 
In the present case, this vanishing theorem holds for irreducible 
conenctions as then the critical points of $\int BF_{A}$ are  of the 
form $F_{A}=B=0$. 
We see that this is precisely the `cofield construction'
employed in \cite{vw} and elaborated upon in \cite{cmr,dm}. 
Once again, the superfield construction proposed in \cite{btmq} provides
a short-cut to constructing actions of this type.

In order
to describe instanton moduli spaces in $d\!=\!4$, all that needs to be done
is to replace the scalar $B$ by a self-dual two-form $B_{+}$. In that
case, there will be a vanishing theorem precisely when the instanton moduli
space is smooth. However, in order to describe theories which localize to
the instanton moduli space under different circumstances, one can also
consider deformations of the condition $F_{+}=0$ by the fields appearing
in the $N_{T}\!=\!2$ $B_{+}$-multiplet
(which will then lead to $F_{+}=0$ under possibly different conditions).
It turns out \cite{vw} that a certain topological twist of
$N\!=\!4$ $d\!=\!4$ Yang-Mills theory (essentially 
constructed first by Yamron \cite{yamron}
and recalled in section 4.2 below) leads to a
particular deformation of this kind and the corresponding vanishing theorems
have been analyzed in detail in \cite{vw}. Taking this modification into
account, it can be verified that the action resulting from (\ref{d24})
agrees term by term with the action constructed in \cite{vw}. 

For both types of theories, the extension to topological gravity or
topological sigma models is immediate. This is also readily seen
in terms of superfields for which the arguments in section 3 imply
the equivalence with the constructions presented in \cite{cmr,dm}.

\subsection{Counting with vs.\ Counting without Signs}

We pause here to comment on one aspect of the above construction 
which we did not state as clearly as we should have in \cite{btmq,sqm}
although it is particularly transparent from the supersymmetric quantum
mechanics point of view. It concerns the question, which was one of the
motivations for introducing the cofield construction in \cite{vw},
`when is one  counting with and when is one counting without signs?'. 
This question arises when the moduli space $\M$ one is localizing onto, the
counterpart of either $X_{V}\ss X$ or $Y\ss X$ in the quantum mechanics 
setting, is not connected. 

In general, counting solutions of
equations, e.g.\ zeros of vector fields or sections of other bundles,
`with relative signs' yields a topological invariant under suitable
compactness conditions. The prime example of this is the Poincar\'e-Hopf
theorem (see (\ref{7})) which expresses the Euler number of a manifold $X$
as the signed sum of zeros of a vector field on $X$. In this case, relative
signs will have to appear not because the total number of solutions is not
a topological invariant (it may be), but simply for the equality with
$\c(X)$ to hold. We will constrast this below with the result (\ref{16})
where $\c(Y)$ is not equated to, say, $\c(X)$, or some other topological
invariant. In that case, relative signs need not (and will not)
appear and under favourable conditions 
the path integral calculates the absolute number of solutions.

Likewise, in its
generalization to non-isloated critical points (\ref{8}), the Euler numbers
of the connected components of the critical point set $X_{V}$ will enter
with relative signs. These signs potentially appear because of the usual
sign ambiguity of fermionic determinants (Pfaffians). While this leaves
undetermined the overall sign, the relative signs, which depend on the 
relative orientations of the connected components $X_{V}^{(k)}$, can be
determined by the spectral flow along trajectories of the (gradient)
vector field connecting two connected components. Thus, $N_{T}\!=\!2$
gauge theory actions based on $S_{X,V}$, in particular the three-dimensional
theory of flat conenctions, count Euler numbers with relative 
signs.\footnote{This played a crucial role in our suggestion \cite{btmq} 
to regard $\c(\M^{(3)})$ as a possible definition for a generalization of
the Casson invariant away from homology spheres.} 

As discussed in detail in \cite{vw}, there are cases where even 
counting solutions without signs may yield a topological invariant,
The conditions for this to occur were phrased in \cite{vw} in terms
of an extended set of fields and equations (the original fields 
supplemented by cofields, the equations for the cofields being - to first
order in the cofields - the linearized original equations) and suitable
vanishing theorems implying that for all solutions to the extended set of
equations the cofields are zero. Because of the constraint on the equations
for the cofields, the determinants fixing the signs of these solutions 
are all positive and one is thus counting solutions of the original equations 
without signs.

However, this is precisely the structure
appearing in the Gauss-Codazzi (Lagrange multiplier) construction (\ref{13})
where
the Lagrange multipliers are the cofields and the Lagrangian is of the form
$\lambda_{a}F^{a} + \ldots $ or $B F_{A} + \ldots $. Explicit evaluation
of the determinants shows that all the fermionic determinants appear in 
pairs (because of the $N_{T}\!=\!2$ structure) so that, for suitable vanishing
theorems, one is counting Euler numbers without signs. This is particularly
clear in the original example, action (\ref{13}), where for a disconnected
submanifold $Y\ss X$ one will simply obtain in (\ref{16}) 
the sum of the Euler numbers $\c(Y^{(k)})$ of its connected components, no
relative signs possibly appearing as the partition function $Z(S_{Y\ss X})$
manifestly just reduces to a sum of the partition functions $Z(S_{Y^{(k)}})$
for actions of the original form (\ref{2}).

Finally, let us point out that the above discussion shows that in addition
to the $d\!=\!3$ theory counting $\c(\M^{(3)})$ with signs (the Casson
invariant when the underlying three-manifold is a homology sphere) there is 
potentially another
one, obtained on using Lagrange multipliers (or the cofield construction)
to localize onto $\M^{(3)}$, which counts them without if appropriate
deformations of the flatness condition and vanishing theorems could be
found. This is just the dimensional reduction of the corresponding theory
for instantons in $d=4$ and its action would schematically be of the
form $\int {\cal B}F_{\cal A}$ where ${\cal B}$ and ${\cal A}$ are $N_{T}=2$
superfields of the type to be discussed in section 3.

\section{$N_{T}\!=\!2$ Superfields and Equivariant Cohomology}

The purpose of this section is to exhibit the relationship between the
equivariant approach to $N_{T}=2$ theories of Dijkgraaf and Moore \cite{dm},
which includes the superfield version of Vafa and Witten \cite{vw}, and 
what at
first sight, appears to be be a completely different superfield
approach employed in \cite{btmq}. 
We will establish the relationship between the various
formulations in the context of gauge theories. It should be
apparent, however, that the equivalence is rather more general.

It is useful to summarise and compare the strategies that are involved. The
approach of Dijkgraaf and Moore is to begin with the so called Weil
model which requires the introduction of 
\be
\sum_{i=1}^{N_{T}} \left( 
\begin{array}{c}
N_{T} \\
i \end{array} \right) = 2^{N_{T}}-1
\ee
`connections' and 
\be
(N_{T}-1)2^{N_{T}} + 1
\ee
independent `curvatures'. The cohomology of interest is basic
cohomology, that is one also demands that the representatives be gauge
invariant (in an appropriate sense). Dijkgraaf and Moore then pass to
the  Cartan model
where the connections are `forgotten' and one deals solely with the
curvatures. The price to be paid is that the cohomology is now
`equivariant', that is it closes only up to gauge transformations
which are parameterised by the curvatures.

The strategy we employed was to consider superspaces of the form
$(x^{\mu},\theta^{m})$ where $m =
1, \dots , N_{T}$, and the $\theta^{m}$ are Grassmann odd space time
scalars and to 
consider a superfield ${\cal A}_{m}$ with the demand that it
transforms as a connection. Given the
connection one can form (linearly dependent) curvatures. The total number of
independent fields is 
\be
N_{T} \times  2^{N_{T}}
\ee
which agrees with the total number of fields in the Weil model of \cite{dm}. 
The usual cohomology arises as supersymmetry transformations (shifts in
the $\theta^{m}$) and requiring `super' gauge invariance. In
order to pass to the Cartan model in this approach one sets
$2^{N_{T}}-1$ connections to zero,
using up all of the `super' gauge invariance that is available.
The left-over terms in the superfield
${\cal A}_{\theta}$ are in fact curvatures. Supersymmetry
transformations change the gauge so, in order to preserve the gauge,
one needs  to do a
compensating super gauge transformation. The combined transformations
are precicely the equivariant exterior derivatives of the Cartan model
of Dijkgraaf and Moore.

As well as the new superfield ${\cal A}_{m}$, there will be the
`physical' or geometric superfield ${\cal A}_{\mu}$ whose zero'th
order component is the gauge field $A_{\mu}$. Depending on the theory
at hand there may also be other multiplier superfields or topological
matter superfields. However, for the general considerations of this
section, we will not have to specify these in any detail. 

The rest of this section is dedicated to filling in the details of
these relationships for $N_{T}=1$ and $N_{T}=2$. We will allude to
the general $N_{T}$ theories but only briefly since the geometry they
encode has still to be determined.

\subsection{Superspace Formulation of $N_{T}=1$ Gauge Theory}

The superspace formulation that we will make use of in this section
was introduced into topological field theory by Horne \cite{horne}.
It was re-interpreted in \cite{vbos} as a method for explaining the
appearance of basic cohomology in topological field theories.

The superspace of interest here is $(x^{\mu}, \theta)$, where
$\theta$ is a Grassman-odd coordinate that is a space-time scalar. One
expands fields in terms of $\theta$, but since $\theta^{2}=0$, the
expansion terminates at the second term. All tensor fields are now
understood to have components also in the $\theta$ direction.

This means that supergauge fields are taken to have a Grassman
component; i.e.\ a supergauge field is a pair $({\cal A}, {\cal
A}_{\theta})$, where ${\cal A}={\cal A}_{\mu}dx^{\mu}$, with the
expansion\footnote{One can shift $\f$ to $\f' = \f - \frac{1}{2}[\xi ,
\xi]$, but the form given in (\ref{conn}) is, as we shall see, canonical.}
\bea
{\cal A}&=& A + \theta \psi \nonumber \\
{\cal A}_{\theta} & =& \xi + \theta \left(\f -\frac{1}{2}[\xi , \xi]
\right)  . \label{conn} 
\eea
Supergauge transformations are
\be
\left( \begin{array}{c}
        {\cal A} \\
        {\cal A}_{\theta} 
\end{array} \right) \rightarrow V^{-1} \left( \begin{array}{c}
        {\cal A} \\
        {\cal A}_{\theta} 
\end{array} \right) V + V^{-1} \left( \begin{array}{c}
        d \\
        \partial_{\theta} 
\end{array} \right)V
\ee
where
\be
V= e^{\theta \lambda} g \label{sgt}
\ee
and $g$ is a conventional gauge transformation.
      
If we construct an action out of our superfields then supersymmetry is
simply the statement that we can shift the grassman coordinate
$\theta$ by $\eps$ without changing anything;
\be
\int d\theta {\cal L}\left( {\cal A}(\theta + \eps)\right) = 
\int d\theta {\cal L}\left( {\cal A}(\theta )\right) .
\ee
Furthermore, we consider an action which is invariant under the
supergauge transformations defined above. In most instances this means
that, up to the inclusion of auxiliary multiplier superfields, the
action will be made out of the superfield curvatures (and covariant
derivatives thereof) associated with the supergauge field. However,
one can make use of a super-version of Chern-Simons theory in three
dimensions as well in which case ${\cal A}$ itself will appear explicitly
in the action.

The supersymmetry transformations then read
\bea
\d {\cal A} &=& {\cal A}(\theta + \eps)- {\cal A}(\theta) \nonumber \\
\d {\cal A}_{\theta} &=& {\cal A}_{\theta}(\theta + \eps)- {\cal
A}_{\theta}(\theta)
\eea
or, in terms of components,
\bea
\d A &=& \psi \nonumber \\
\d \psi &=& 0 \nonumber \\
\d \xi &=& \f - \frac{1}{2}[\xi , \xi] \nonumber \\
\d \f & = & [ \f , \xi ]. \label{shift}
\eea

This description of the theory is highly redundant. One can express
the superfield ${\cal A}_{\theta}$ as
\be
{\cal A}_{\theta} = U^{-1}{\cal A}_{\theta}^{0}U +
U^{-1}\partial_{\theta}U  \label{rep1}
\ee
where
\be
{\cal A}_{\theta}^{0} = \theta \f , \label{nfield}
\ee
\be
U(\xi)= \exp{(\theta \xi)} .
\ee
The virtue of writing things in this way is that we see immediately that {\em
all} of the `super' part of the supergauge transformations is taken
up in the field $\xi$. This means that in a supergauge invariant action
the field $\xi$ will not appear since it is pure gauge. One sees this
most clearly by considering the
action of the super-part of a supergauge
transformation. Under such a gauge transformation (i.e. $g=I$ in
(\ref{sgt})) from (\ref{rep1}) we see that
\be
U(\xi) \rightarrow U(\xi)U(\lambda) = U(\xi + \la ) \label{grassym}
\ee
which corresponds to a shift of $\xi$, while $\f$ is invariant
\bea
\xi & \rightarrow & \xi + \lambda \nonumber \\
\f & \rightarrow & \f . \label{sgt1}
\eea

The transformation (\ref{grassym}) means that we can choose a gauge
where there is no $\xi$ component for ${\cal A}_{\theta}$. This gauge
is achieved by taking
\be
\lambda = -\xi
\ee
since in this case $U(\xi)U(-\xi )=1$.

We should always be able to work with fields of the form
(\ref{nfield}) at the cost of only allowing conventional gauge
transformations. This would be correct if we were not interested in
supersymmetry as well. The problem is that under a supersymmetry
transformation 
\bea
{\cal A}_{\theta}(\theta) \rightarrow & & {\cal A}_{\theta}(\theta +
\eps)  \nonumber
\\
&=& \eps \f + \theta \f
\eea
so that a $\xi$ term ($= \eps \f$) reappears. However, we know that
by a gauge transformation we can eliminate {\em any} $\xi$ term. So
the strategy is to follow the shift in $\theta$ by a gauge
transformation that puts the superfield back into the form where there
is no $\xi$ component. This is easy to do and it is straightforward to
determine the gauge transformation that is required (it is
$\exp{(-\theta\eps\f )}$) though in fact we will not need its explicit form
in the following. 

A consequence of this is that the action, with the reduced field
content (that is with $\xi=0$), will only be invariant under
the combined shift and supergauge transformation (as well as
conventional gauge transformations). We will deduce the appropriate
symmetry by a slightly indirect method that, nevertheless, will be
useful when we come to comparing with the work of Dijkgraaf and Moore.
We introduce the field strengths that are associated with the
connections (\ref{conn}) (we have dropped the superscript $0$)
\bea
{\cal F}_{A} &=& d{\cal A} + {\cal A}^{2} \nonumber \\
             & = & F_{A} + \theta d_{A}\psi \nonumber \\
{\cal F}_{\theta } &=& \partial_{\theta}{\cal A} -d {\cal A}_{\theta} 
- [{\cal A}, {\cal A}_{\theta}] \nonumber \\
& =& \partial_{\theta}{\cal A} -d_{{\cal A}}{\cal A}_{\theta}
\nonumber \\ 
                  &=& \psi - \theta d_{A}\f \nonumber \\
{\cal F}_{\theta \theta} &=&2\partial_{\theta}{\cal A}_{\theta} +
{\cal A}_{\theta}^{2} \nonumber \\
         & =& 2\f . \label{curv}
\eea
Notice that $\f$ from (\ref{curv}), though it appears in ${\cal
A}_{\theta}$, is indeed a curvature\footnote{We have
been continually referring to $\xi$ as a connection in the superfield
approach. The reason for this nomenclature is that if one calculates
curvatures prior to gauge fixing then $\xi$, like $A$, never appears
as the lowest component of a curvature superfield. For example ${\cal
F}_{\theta} = \psi -d_{A}\xi + O(\theta ) $. }.

To deduce the transformations, we note that for all the field
strengths, under the combined shift and gauge transformation we have
\be
{\cal F}(\theta) \rightarrow U^{-1}(\theta, \eps) {\cal F}(\theta +
\eps) U(\theta, \eps) , \label{trans}
\ee
where $U(\theta , \eps)$ is the gauge transformation which puts the
superfield back into the form (\ref{nfield}) and necessarily $U(0,
\eps) =I =  U(\theta , 0)$. 
Now we will argue that one can safely ignore 
$U(\theta , \eps )$ in determining the transformation properties of
the fields. The argument is as follows: we wish to determine the
transformation of the fields that appear at $O(1)$ in the field
strengths, that is $\psi$ and $\f$ but, since $U= I + O(\theta)$, $U$ can
only contribute to terms of order $\theta$ so we may safely for
present purposes set it to unity. We therefore, when dealing with the
curvatures, only need to consider shifts in $\theta$. From
(\ref{curv}) we easily deduce that the topological transformation is
\bea
Q A &=&  \psi \nonumber \\
Q \psi &=& - d_{A} \f \nonumber \\
Q \f &=& 0 \label{trans1}
\eea
(we obtained the $A$ transformation using the same logic but
applied to ${\cal A}$). Let us repeat that we have derived these
transformations without having to know the explicit form of $U(\theta ,
\eps)$.

\subsection{Weil and Cartan Models}

We will now relate the above construction to something that appears
in the mathematics literature. For $G$ a Lie group and $\lg$ its
algebra, we define
the Weil algebra ${\cal W}(\lg)$ by
\be
{\cal W}(\lg) = \Lambda (\lg^{*}) \otimes S(\lg^{*})
\ee
which is the tensor product of the exterior algebra and the symmetric
algebra of the dual $\lg^{*}$. Denote the generator of $\Lambda
(\lg^{*})$ by $\xi$ with degree 1 (it is a connection) and the
generator of $S(\lg^{*})$ of degree 2 (a field strength) by $\f$. On
the Weil algebra one can define a covariant derivative $d_{W} + \xi$
where $d_{W}$ is a derivation on ${\cal W}$. The relationship between
the various objects is
\be
\f = \frac{1}{2}[d_{W} + \xi , d_{W} + \xi] , \;\;\; d_{W} \f + [\xi ,
\f ] =0 \label{weil1}
\ee
the second equation being the Bianchi identity. These are rewritten
as
\bea
d_{W}\xi &=& \f - \frac{1}{2}[\xi , \xi] \nonumber \\
d_{W} \f & = & [\f , \xi ] . \label{weil2}
\eea

The cohomology of the Weil complex is trivial since, by shifting
$\f' = \f - [\xi , \xi ]/2$, one finds
\be
d_{W}\xi = \f' , \;\;\; d_{W}\f' =0 .
\ee
The cohomology of interest is actually the so called `basic'
cohomology. The interior derivative $i_{a}$ is defined by
\be
i_{a}\xi^{b} = \delta_{a}^{b}, \;\;\; i_{a}\f^{b} = 0 , \label{inner}
\ee
A form is called basic if it is horizontal and gauge invariant, i.e.
it satisfies
\be
i_{a}\om =0, \;\;\; {\cal L}_{a}\om =0 
\ee
where
\be
{\cal L}_{a} = [d_{W} , i_{a} ] .
\ee

In this context it is useful to pass to the Cartan model.
Formally, this amounts to `forgetting' $\xi$. This means that one
sends $\xi \mapsto 0$ and considers this as a map of ${\cal
W}(\lg^{*}) \rightarrow S(\lg^{*})$ which induces an algebra
isomorphism
\be
{\cal W}(\lg^{*})_{\mbox{basic}} \equiv S(\lg^{*})^{G}
\ee
where the superscript $G$ means the $G$-invariant subalgebra. This is
an isomorphism of differential (graded) algebras when one defines the
derivation
\be
d_{C}\f =0
\ee 
on $S(\lg^{*})^{G}$. 

Let us generalise a little. Consider the algebra ${\cal W}(\lg^{*})
\otimes \Omega^{*}(M)$, where $M$ is some manifold on which $G$ acts.
Define the derivation on this algebra by
\be
d^{W} = d_{W}\otimes 1 + 1 \otimes d .
\ee
The basic forms in this case are defined by, for $\omega \in {\cal
W}(\lg^{*}) \otimes \Omega^{*}(M)$,
\be
i_{a}\om = {\cal L}_{a}\om =0 .
\ee
To pass to the Cartan model we have to ask what the effect of the
`forgetting $\xi$' is on the extended space. The result is that one
defines the Cartan derivation in this case as
\be
d^{C} =  1 \otimes d - \f^{a}\otimes i_{a} .
\ee

\subsection{Equivalence of the Weil, Cartan and Superfield Models}

We begin with the trivial observation that the field content, $(\xi ,
\f )$, of the Weil and the superfield models are the same. Indeed the
fields have the same interpretation, $\xi$ is a connection while $\f$
is a curvature. Secondly, the action of the exterior derivative
$d_{W}$ in the Weil model (\ref{weil2}) agrees with the supersymmetric
shift operation (\ref{shift}), so we conclude that
\be
\d = d_{W}
\ee
In order to pass to the basic cohomology in the Weil model one needs
an inner derivation. By comparing
(\ref{inner}) and (\ref{sgt1}) we see that what are called
super gauge transformations $\exp{(\theta \la)}$ correspond precicely
to the derivations $\la^{a}i_{a}$. In other words basic forms are, in the
superfield language, gauge invariant forms.

The isomorphism between the Weil and the Cartan models, as graded
differential algebras, only works
when one restricts ones attention to basic forms in the Weil model
since on setting $\xi=0$ we would deduce from (\ref{weil1}) that $\f
=0$. In
the time honoured tradition of gauge theory we either work
equivariantly or we gauge fix. Working equivariantly in the Weil
model means working with basic forms while gauge fixing means working
with the Cartan model. The gauge we would like to choose is $\xi=0$. We
saw above that if we do this then we would be led to
\be
\d 0= d_{W}0 =\f
\ee
and that a way out is to supplement the derivation $\d =\partial / \partial
\theta$ with a gauge transformation, and we called the new derivation
$Q$, so that 
\be
Q {\cal A}_{\theta} = 0
\ee
and we identify, on $S(\lg^{*})^{G}$,
\be
d_{C} = Q ,
\ee
(and setting $\xi$ to zero is now fine since the derivation equation is
$d_{C}\xi =0$). 
Somewhat more generally the action of the group $G$ is understood to be
simultaneously an action on both factors of ${\cal W}(\lg^{*})
\otimes \Omega^{*}(M)$. $d_{C}$ was obtained by supplementing $d_{W}$
with a gauge transformation, i.e. $d_{C} = d_{W} - \f^{a}i_{a}$. A
natural derivation is that obtained on gauge transforming both
differentials. So we are led to
\be
d^{C} = d_{C} \otimes 1 + 1 \otimes d - \f^{a}\otimes i_{a}
\ee
which is the standard differential in the Cartan model (the first term
gives zero).

Now, what this translates to for gauge theories, in the `geometric'
sector is, since the group that acts is the group of gauge
transformations, 
\be
d^{C}A=0 , \;\;\; d^{C}\psi = d_{A}\f , \;\;\; d^{C}\f =0 .
\ee
These agree with the action of $Q$, but this comes as no surprise
since we have seen in the previous section that this is
{\em exactly} how the BRST operator arises in the $N_{T}=1$ gauge
theory. In fact the relationship between
$Q$ and $d^{C}$ is rather transparent when one looks at the curvature
${\cal F}_{\theta}$ (\ref{curv}) which we rewrite as
\be
{\cal F}_{\theta} = \left( \frac{\partial}{\partial \theta } - i_{{\cal
A}_{\theta}} \right) {\cal A} .
\ee

It is also worthwhile exhibiting the relationship by determining $U$,
since this approach will facilitate the comparison for other values
of $N_{T}$. We have commented before that
\be
U= e^{-\theta \eps \f} = e^{-\theta {\cal A}_{\theta}(\eps)}
\ee
so that to first order in $\eps$ the definition of $Q$ is: perform a
translation, i.e.\ act by $1 \otimes d$, and follow that by a gauge
transformation $U = 1 - \theta \eps \f$, i.e.\ act by $-\f^{a}\otimes
i_{a}$. 

\noindent\underline{{\bf Remarks:}} 

\begin{enumerate}
\item
The relationship with the paper of
Dijkgraaf and Moore is clear. Their connection $\om$ is what we have
called $\xi$ and we agree on the symbol for the curvature $\f$. 
\item
In the superfield approach both the connection $\xi$ and the
curvature $\f$ appear in the same superfield. This happens because,
from this point of view, curvatures are not independent of
connections. 
\item
One does not even need to know the transformations (\ref{trans1})
to construct an invariant action. On using the curvature superfields we are
assured that we will have an action which is $Q$ invariant and $Q$
exact. Furthermore, without knowing what $Q$ is we know that $Q^{2}=
{\cal L}_{\la}$, for some $\la$, by construction. It is this level of
generality which makes the superfield approach easy to use.

\item There is also a well known procedure for arriving at topological
observables. One constructs characteristic classes of the super
curvatures. 

\item In order to determine the transformations of multiplier superfields,
once $\xi$ has been set to zero, one uses the same logic that we
employed on the curvatures, but this time on covariant derivatives of
the multiplier fields.
\end{enumerate}

\subsection{Superfield Formulation of $N_{T}=2$ Gauge Theory}

The superspace in question is that spanned by two Grassmann
coordinates which we denote by $\theta^{m}$, $m=1,2$. Some notation is
required. Let
\be
\theta^{2} = \frac{1}{2}\eps_{mn}\theta^{m}\theta^{n} , \;\;\;
\eps_{12}= 1 ,
\ee
then we find that
\be
\theta^{m}\theta^{n} = \eps^{mn}\theta^{2} , \;\;\; \eps^{12}=1 .
\ee
There are two Grassmann components of the gauge field which are
denoted by ${\cal A}_{m}$. As for the $N_{T}=1$ theory a meaningful
way to represent the super gauge field is as follows
\be
{\cal A}_{m} = U^{-1}{\cal A}_{m}^{0}U + U^{-1} \partial_{m} U \label{rep2}
\ee
where
\be
{\cal A}_{m}^{0} = \theta^{n}\f_{nm} + \theta^{2}\eta_{m} ,
\ee
\be
U= \exp{\left( \theta^{m}\xi_{m} +\theta^{2}\rho\right) } ,
\ee
and $\f_{mn} =\f_{nm}$.
The $\xi_{n}$ and $\rho$ are connections and
the $\f_{nm}$ and $\eta_{m}$ are (generalised) curvatures. We note
that the connections match with the fields $\om_{\pm}$ and $\Omega$ of
Dijkgraaf and Moore, as do the curvatures which bear the same name
(cf.\ \cite[eq (3.16)]{dm}).

The Grassmann part of the gauge group has three independent variables
\be
U = \exp{\left(\theta^{m}\la_{m} + \theta^{2}\sigma\right)} 
= 1 + \theta^{m}\la_{m} + \theta^{2}(\sigma - \lambda^{2}) ,
\ee
with $\la^{2} = \frac{1}{2}\eps^{mn}\la_{m}\la_{n}$. Under such a super-gauge
transformation we have
\be
{\cal A}_{m} \rightarrow U^{-1}{\cal A}_{m}U + U^{-1} \partial_{m} U ,
\ee
or
\be
\exp {\left(\theta^{m}\xi_{m} + \theta^{2}\rho\right)} \rightarrow
\exp {\left(\theta^{m}\xi_{m} + \theta^{2}\rho\right)}
\exp {\left(\theta^{m}\la_{m} + \theta^{2}\sigma\right)}
\ee
from which we deduce that
\bea
\d  \xi_{m} &=&  \la_{m} \nonumber \\
\d \phi_{nm}
&=& 0 \nonumber \\
\d \rho &=& \sigma + \frac{1}{2} \eps^{nm}[ \la_{n} , \xi_{m}] 
\nonumber \\
\d \eta_{m} &=& 0 .
\eea
Since the gauge field ${\cal A}_{m}$ is the gauge transform of ${\cal
A}_{m}^{0} $ we need only calculate the curvature tensors for ${\cal
A}_{m}^{0}$. We will drop the superscript. The usual gauge field $A$
is the first component of a superfield
\be
{\cal A} = A + \theta^{m}\psi_{m} + \theta^{2}B .
\ee
The general definition of
the curvature tensor for a superspace with $N_{T}$ scalar grassmann
coordinates in $d$ dimensions is
\be
{\cal F}_{MN} = \partial_{M}{\cal A}_{N} - [MN] \partial_{N}{\cal
A}_{M} + [{\cal A}_{M} , {\cal A}_{N} ]
\ee
with $M,N = (\mu , m), (\nu , n)$ where $\mu , \nu = 1, \dots , d$ and
$ m , n= 1, \dots , N_{T}$. The symbol $[MN] = \pm 1$, is $-1$ only if
$M=m$ and $N=n$ simultaneously.

One finds that, with $N_{T}=2$,
\be
{\cal F}_{mn} = 2\f_{mn} + \theta^{p}(\eps_{mp}\eta_{n}+
\eps_{np}\eta_{m}) + O(\theta^{2})
\ee
and
\be
\eps^{mp}[D_{p}, {\cal F}_{mn}] =
3\eta_{n}-3\theta^{m}\eps^{rs}[\f_{rm},\f_{sn}] 
+ O(\theta^{2})
\ee
\be
{\cal F}_{m } =  \psi_{m} + \eps_{m n}\theta^{n} B -
\theta^{n}d_{A}\f_{nm} + O(\theta^{2}) 
\label{calfm}
\ee
\be
\eps^{mn}D_{n}{\cal F}_{m} = 2 B + \theta^{m}d_{A}\eta_{m} + \eps^{mn}
\theta^{p} 
[ \psi_{n} , \f_{pn}]  + O(\theta^{2})  .
\ee

Using the fact that curvatures transform homogeneously and that $U$ is
necessarily of order $\theta$ we deduce from the curvatures above that
\bea
Q_{n}A & =& \psi_{n} \nonumber \\
Q_{n}\psi_{m} & =& \eps_{mn} B - d_{A}\f_{mn} \nonumber \\
Q_{n}B &=& \frac{1}{2}d_{A}\eta_{n} + \frac{1}{2}\eps^{mp}[ \psi_{p} ,
\f_{mn} ] \nonumber \\ 
Q_{n}\f_{mp} & =& \frac{1}{2}\eps_{mn}\eta_{p} +
\frac{1}{2}\eps_{pn}\eta_{m} \nonumber \\
Q_{n}\eta_{m} & = & \eps^{pq}[ \f_{nq} , \f_{mp} ] .
\eea

These transformations agree with those of \cite{vw} and \cite{dm} 
(taking into account that their $\eta_{m}$ and ours differ by a 
factor of two). In \cite{btmq} we gave a different method for
determining these transformations which amounts essentially to 
fixing $U$ directly. We will follow a similar course in the next 
section when we come to comparing with \cite{dm}.

\subsection{Equivalence of the Weil, Cartan and Superfield Models for
$N_{T}=2$}

Rather than reviewing each model seperately we can proceed directly to
a comparison. The Weil model \`{a} la Dijkgraaf and Moore has three
connections $(\xi_{m},
\rho)$ and field strengths $(\f_{nm} , \eta_{m})$. These fields are
the components of ${\cal A}_{m}$ in the superfield language.
As for the $N_{T}=1$ theories we would like to pass from the Weil
model to the Cartan, and as before we do this by gauge fixing the
`connections' $(\xi_{m},\rho)$ to zero (as we have done for the
superfields). Next we identify the derivations. In the equivariant
cohomology of Dijkgraaf and Moore there are derivations $i^{m}$ and
$I$ which satisfy the algebra 
\be
[i^{m}(V), i^{n}(W) ] = \eps^{pm} I \left( [ W , V ] \right) \label{alg}
\ee
with other (anti-) commutators being zero. They act on the connections
by 
\be
i_{ a}^{m}\xi_{n}^{b} = \d^{m}_{n}\d^{b}_{a} ,\;\;\; i^{m}_{ a}\rho^{b}
= \frac{1}{2}\eps^{m n} [ T_{a} , \xi_{n} ]^{b} 
\ee
and
\be
I_{a}\xi^{b}_{m} = 0 , \;\;\; I_{a}\rho^{b} = \d^{b}_{a} .
\ee
In our model these correspond to the possible `super' gauge
transformations. By inspecting the supersymmetry transformations
(\ref{sgt1}) we see that the $\lambda_{m}$ part
of the gauge transformation corresponds to $i^{m}$ while in the
$\sigma$ part one recognises the action of $I$. The algebra that they
satisfy is easily deduced by composing two superfield gauge
transformations whose exponents depend only $\theta^{m}$ and writing
the product as a single superfield gauge transformation which has in
the exponent a term proportional to $\theta^{2}$. The algebra of
course agrees with (\ref{alg}).

Now the definition of the Cartan derivations on the physical fields,
following \cite{dm}, is
\be
d^{C}_{m} = d^{W}_{m} + \f^{a}_{mn}i^{n}(V_{a}) + \eta^{a}_{m}I(V_{a})
\ee
which agrees with the action of $Q_{m}$. However, it is again enlightning 
to see this
in terms of $U$. To determine $U$, we note that what we would like is
the solution to the equations
\bea
{\cal A}'_{m}|_{\theta =0} & = & {\cal A}_{m}(\eps) +
\partial_{m}U(\theta , \eps)|_{\theta =0} = 0 \non
\eps^{mn}\partial_{n}{\cal A}'_{m}|_{\theta =0} & = & \eps^{mn} \left( 
\partial_{m}U(\theta , \eps)|_{\theta =0} \partial_{n}U(\theta ,
\eps)|_{\theta =0}\right.\non&&  +\left.
\partial_{n} {\cal A}_{m}(\eps) + \partial_{n}\partial_{m}U(\theta ,
\eps)|_{\theta =0} \right) = 0 
\eea
The solution to these equations is
\be
U = \exp{ \left( - \theta^{m}{\cal A}_{m}(\eps) -\frac{ \theta^{2}}{2}
\eps^{mn}\partial_{n} {\cal A}_{m}(\eps) \right) } . \label{gderivs}
\ee
We have already identified the $\theta$ term of a gauge transformation as the
coefficient of a derivation $i^{m}$ and the $\theta^{2}$ term as the
coefficient of the derivation $I$ so that by inspection of
(\ref{gderivs}) and to first order in $\eps$ we have that $U$ acts by
\be
-\eps^{m}\left[ \f^{a}_{mn}i^{n}_{a} - \frac{1}{2}\eta^{a}_{m}I_{a}
\right]
\ee
once more in complete agreement (up to the ubiquitous factor of $2$ in
front of $\eta$) with \cite{dm}.

Most of the remarks that we made for the $N_{T}=1$ models hold here as
well with slight modification. In particular, 
any action made out of the superfield curvatures will be $\eps^{mn}
Q_{m} Q_{n}$ exact.

One peculiarity of three dimensions is that there
the `smallest' topological gauge theory appears to
have $N_{T}=2$. One could construct, formally, an $N_{T}=1$ super-BF
theory, but it is easy to see that this action agrees with that which
one obtains from the $N_{T}=2$ Chern-Simons action. This also explains
why the partition function comes in with signs. We will establish a more
general result (for arbitrary $N_{T}$) in the next section.

In summary, we have seen that the algebraic 
constructions of Vafa and Witten
\cite{vw} and of Dijkgraaf and Moore \cite{dm} are indeed identical to
the geometric superfield approach that we introduced in \cite{btmq}, the
two points of view being complementary. A 
variant on this theme preserving the underlying $sp(2 ,\RR)$
symmetry and which does not set the connections to zero (corresponding
to an $N_{T}=2$ version of the so-called BRST model \cite{kalkman} of
equivariant cohomology) can be found in \cite{gtlec}.

\subsection{Arbitrary $N_{T}$}

As Dijkgraaf and Moore briefly consider the extension of their formalism to
arbitrary $N_{T}$, let us indicate how this can be done in terms of
superfields. 

An arbitrary $N_{T}$ gauge field has as an expansion
\be
{\cal A}_{m} = \sum_{i=0}^{N_{T}} \theta^{j_{1}} \dots
\theta^{j_{i}}\Phi_{j_{1} \dots j_{i} ,m}
\ee
the components of which are products of a totally antisymmetric tensor
in $i$ labels and the fundamental vector $N_{T}$. Such a product can always
be written as the direct sum of a $i+1$ totally antisymmetric tensor
and a tensor of mixed symmetry (with Young tableau consisting of two
columns, the first has $i$ rows the second just one row).

Gauge parameters have an expansion in terms of forms (i.e.\ totally
antisymmetric tensors), so that 
\be
U = e^{\Lambda}
\ee
with
\be
\Lambda = \sum_{i=1}^{N_{T}}\theta^{j_{1}} \dots
\theta^{j_{i}}\rho_{j_{1} \dots j_{i}}
\ee

We conclude then that any ${\cal A}_{m}$ can be written as the gauge
transform of 
\be
{\cal A}_{m}^{0} = \sum_{i=0}^{N_{T}} \theta^{j_{1}} \dots
\theta^{j_{i}}\Phi_{j_{1} \dots j_{i} m} ,
\ee
where
\be
\Phi_{j_{1} \dots j_{i} m}
\ee
is the tensor of mixed symmetry.

How many derivations do we have in this model? The answer is that the
number of derivations matches the number of `super' gauge parameters,
so we have the derivations
\be
i_{m} , \, i_{mn} , \, \dots , \eps_{m_{1}, m_{2}, \dots m_{N_{T}}}I .
\ee
One can, once more, determine the algebra of these by
expanding the superfields. 

If one wishes to make contact with the work of \cite{dm} then one can
straightforwardly determine the appropriate $U$, bearing in mind that
one requires it only to first order in $\eps$ (higher orders tell one
about successive BRST transformations). One now wants to solve
\be
{\cal A}'_{m}|_{\mbox{tas}} = \left( U^{-1}{\cal A}_{m}(\theta +
\eps)U + U^{-1}\partial_{m}U \right)|_{\mbox{tas}} =0 \label{ueq}
\ee
where tas stands for the totally antisymmetric tensor part. Writing
$U$ to first order in $\eps$, as
\be
U = 1 + \eps^{m}\Lambda_{m}(\theta) + \dots
\ee
(\ref{ueq}) becomes
\be
[ {\cal A}_{m}(\theta) , \eps^{n}\Lambda_{n} ]|_{\mbox{tas}} +
\eps^{n}\partial_{n} {\cal A}_{m} |_{\mbox{tas}} -
\eps^{n}\partial_{m}\Lambda_{n} |_{\mbox{tas}} = 0 ,
\ee
which can be solved order by order in $\theta$.

While to date there has been no application of these types of
topological field theories
with $N_{T} \geq 3$ and their meaning remains unclear,
in three dimensions a possible action for $N_{T} = 2n$ is 
the super Chern-Simons functional. This does not work for
$N_{T}= 2n+ 1$ since the top component of ${\cal A}$ is Grassmann odd
in that case. 

We will now show that ${\cal B}{\cal F}_{{\cal A}}$ theory in 3 
dimensions with $N_{T}=2n-1$ is the same as the superfield Chern-Simons 
action for $N_{T}=2n$, thus eliminating the possibility of odd numbers of 
supersymmetries in this dimension (unless, of course, one breaks the 
symmetry down by hand). Let $N_{T}=2n-1$ and let the Grassmann 
coordinates be $\theta^{m}$ with $m = 1 , \dots , N_{T}$. For the action 
to make sense the lowest component of ${\cal B}$ must be Grassmann odd. 
Introduce a new Grassmann variable $\theta = \theta^{N_{T}+1}$ and let
\be
{\cal A}(\theta^{m} , \theta) = {\cal A}(\theta^{m}) + \theta {\cal 
B}(\theta^{m}) \label{n+1}
\ee
This is the most general expansion of a gauge superfield in $\theta$, and 
so this is the most general expression of a gauge superfield in a 
superspace with $2n$ Grassmann coordinates. The Chern-Simons action is, 
\be
\prod_{i=1}^{N_{T}+1} \int d\theta^{i} S_{CS}\left( {\cal A}(\theta^{m} , 
\theta) \right) = \prod_{i=1}^{N_{T}} \int d\theta^{i} {\cal B}{\cal 
F}_{{\cal A}} .
\ee
Therefore there is an increase in the amount of supersymmetry in the 
theory.

The even $N_{T}$ super-BF theories in 3 dimensions, on the other hand,
can also be 
realised as super Chern-Simons theories with (a priori) the same $N_{T}$, 
but with the gauge group being $IG$, the tangent bundle group of $G$. 
The most well known example of this is the case $N_{T}=0$ where the
fields $A$ and $B$ of the $G$-BF theory can be combined into a connection
$A+B$ for $IG$. 

\section{Other $N_{T}\!=\!2$ Topological Gauge Theories}

In this section, we will briefly describe two other topological gauge
theories with an extended $N_{T}\!=\!2$ topological superysmmetry  which
do not fall into the pattern of $N_{T}\!=\!2$ theories described above. 
It is well known that extended supersymmetries can arise when $N_{T}\!=\!1$
theories are formulated on manifolds with reduced holonomy groups
(e.g.\ Donaldson-Witten theory on K\"ahler manifolds). The extended 
supersymmetries we will encounter below bear some resemblance to this,
resulting however, roughly speaking, from a complexification of the 
space of fields
and not from, say, a K\"ahler stucture on the space-time manifold.

As a preparation, in section 4.1  we recall the structure of super-BF
theories in $d\!=\!3,4$ \cite{bbt,bbrt} and construct the quantum action
using a minimal field content not nearly filling out the Batalin-Vilkovisky
triangles by exploiting the topological supersymmetry present in these
models. 

In section 4.2, we show that the $N_{T}\!=\!2$ theory constructed by Marcus 
\cite{nm}, the B-twist of $N\!=\!4$ $d\!=\!4$ Yang-Mills theory,
can be regarded
as a deformation of the four-dimensional $N_{T}\!=\!1$ super-BF theory
\cite{bbt,bbrt}, but we refrain from a detailed discussion of that model
as it is not clear that the study of moduli spaces of flat connections
in $d\!=\!4$ is particularly meaningful. 

In section 4.3 we describe a novel twist of 
$N\!=\!4$ $d\!=\!3$ Yang-Mills theory.
This twist uses the internal $SU(2)$ arising in the reduction of $N\!=\!1$ 
$d\!=\!6$ Yang-Mills theory to $d\!=\!3$ and leads to some unusual features
which we briefly describe. 

\subsection{Simplified Quantization of Super-BF Theories}

Super-BF theories are cohomological gauge theories of flat connections. They
can be defined in any dimension $d$, and the action typically takes the form
$\int BF_{A}$  plus superpartners plus gauge fixing terms, 
where $F_{A}$ is the curvature of the connection $A$, the mutiplier field
$B$ is a $(d-2)$-form in the adjoint of the gauge group and a trace is 
understood. The $N_{T}\!=\!1$ superpartners are $QA=\p$, $Q\c =B$, 
which will always be supplemented by an anti-ghost multiplier pair
$(\fb,\e)$ for $\p$ with $Q\fb=\e$, $Q\e=[\fb,\f]$
so that the `classical' action can be taken to be
\bea
S_{c}^{(d)}&=&Q\int\c F_{A} +\fb d_{A}*\p \non
          &=&\int BF_{A} + (-1)^{d}\c d_{A}\p + \e d_{A}*\p + \fb [\p,*\p]
              -\fb d_{A}*d_{A}\f
\eea
As reviewed in detail in \cite{bbrt}, the universal 
terms arising from the variation
of the second term encapsulate concisely the geometry of the universal
bundle over the space of gauge orbits - restricted to the moduli space of
interst by the terms arising from the variation of the first term.
 
This action has the tower $\d B = d_{A}B^{(1)}$, $\d B^{(1)}=d_{A}B^{(2)}$,
\ldots, of on-shell reducible symmetries and its quantization was analysed 
in detail in \cite{bbt,bbrt} using the Batalin-Vilkovisy procedure. 
In the simplest non-trivial case
$d\!=\!3$ this leads to the following field content and action: 

In addition to the universal geometric sector $(A,\p,\f,\fb,\e)$ and the 
multiplier pair $(\c,B)$ one has scalar ghost-antighost-multiplier triplets
$(\sigma,\bar{\sigma},\pi)$ and 
$(\Sigma,\bar{\Sigma},\Pi)$ for $\c$ and $B$ respectively.
The off-shell equivariantly nilpotent $Q$-transformations are
\bea
QA = \p && Q\p = d_{A}\f \non
Q\f =0 &&\non
Q\fb = \e && Q\e = [\fb,\f]  \non
Q\c = B + d_{A}\sigma && QB = d_{A}\Sigma + [\c,\f] + [\p,\sigma] \non
Q\sigma = \Sigma&& Q\Sigma = [\sigma,\f]  \non
Q\bar{\sigma} = \pi && Q\pi= [\bar{\sigma},\f] \non
Q\bar{\Sigma} = \Pi && Q\Pi= [\bar{\Sigma},\f] 
\eea
and the complete quantum action can be chosen to be
\be
S_{q}^{(3)} = Q\int \c F_{A}+\fb d_{A}*\p + \bar{\sigma}d_{A}*\c +
\bar{\Sigma}d_{A}*B\;\;.\label{sbfact1}
\ee

However, due to the $Q$-pairing there is clearly some redundancy in this
description as e.g.\ the variation of the third term all by itself already
provides a gauge fixing for $\c$ and $B$. Moreover, by a field redefinition
of $B$ the ghosts $\sigma$ and $\Sigma$ can be eliminated from the picture.

This reasoning suggests an alternative procedure, bypassing the BV algorithm.
Instead of two ghost-triplets one introduces just one antighost-multiplier
pair $(\bar{\sigma},\pi)$ which we will now call $(u,\eb)$ 
and postulates the equivariantly nilpotent algebra
\bea
QA = \p && Q\p = d_{A}\f \non
Q\f =0&& \non
Q\fb = \e && Q\e = [\fb,\f]  \non
Q\c = B && QB = [\c,\f]  \non
Qu = \eb& & Q\eb= [u,\f] 
\label{qred}
\eea
The corresponding action for this reduced field content,
\be
S_{q}^{(3),red}=Q\int\c F_{A} + \fb d_{A}*\p + ud_{A}*\c\;\;,
\label{sbfact2}
\ee
can readily be seen to be equivalent to (\ref{sbfact1}) by integrating out
the superfluous fields.\footnote{To be precise, this holds modulo harmonic 
modes of the ghost-fields which have to be dealt with in some manner in the 
original action anyway.} 

Note that the field content suggests a 
discrete symmetry exchanging $\e$ and $\eb$ and $\c$ and $\p$, so that 
there should actually be a (presently hidden) 
$N_{T}\!=\!2$ symmetry in this theory. This is indeed the case. In fact,
this theory is precisely the $N_{T}\!=\!2$ theory (\ref{qmsbf3})
calculating the Euler 
characteristic of the moduli space of flat connections which we have
discussed in detail in \cite{btmq}. The reduced
field content is that one obtains after eliminating the super-gauge
transformation. In order to obtain a manifestly $N_{T}\!=\!2$ symmetric
action, one needs to add to (\ref{sbfact2}) 
a term of the form $Q\int \c *B$, thus 
regularizing the harmonic modes of $B$, and then integrate out $B$.

As we will recall below, this is also one of the actions one
can obtain by twisting the $N\!=\!4$ $d\!=\!3$ Yang-Mills theory. From this
point of view, the $N_{T}\!=\!2$ symmetry is manifest as the fermions 
$(\e,\eb)$ and $(\c,\p)$ then appear as doublets of the $SU(2)$ 
ghost-number symmetry of the theory. As $SU(2)$ is simple, this ghost
number is not anomalous and the partition function is non-zero. 

The four-dimensional super-BF theory has been discussed at length in
\cite{bbt,bbrt}. Its quantization is non-trivial as the $\chi$-symmetry
is now on-shell reducible, leading to the ghost-for-ghost phenomenon
and cubic ghost interaction terms. Following the BV algorithm, one 
obtains an impressive field content consisting of a ghost and ghost for
ghost for both $\c$ and $B$ and three antighost-multiplier pairs each,
in addition to the fields $(A,\p,\f,\fb,\e)$. The corresponding quantum
action, with an off-shell equivariantly nilpotent $Q$-symmetry, is 
spelled out in \cite[eq.\ (5.133)]{bbrt}. 

In this case, the procedure outlined above for arriving at a reduced field 
content leads to a significant simplification. The transformations on the
fields $(A,\p,\f,\fb,\e,\c,B)$ remain as in (\ref{qred}).
As $\chi$ and $B$ are
now two-forms, instead of the pair of scalars $(u,\eb)$ we introduce
a pair of one-forms $(V,\pb)$ as well as the ghost-for-ghosts $(\eb,u)$
(to account for the symmetries of $(V,\p)$) with
\bea
 QV=\pb&& Q\pb=[V,\f]\non
 Q\eb=u && Q u=[\eb,\f]\;\;.
\label{qred2}
\eea
The quantum action is now taken to be
\be
S_{q}^{(4),red}=Q\int\c F_{A} -\fb d_{A}*\p + V d_{A}*\c - \eb d_{A}*V
\label{sbf4act}
\ee
and clearly all the local symmetries of the action apart from ordinary
gauge invariance have been gauge-fixed. The appearance of both $\p$ and
$\pb$ once again suggests the presence of a hidden $N_{T}\!=\!2$ symmetry,
although of a different kind this time as both $\p$ and $\pb$ have
ghost-number one. It is the aim of the next section to expose this symmetry
and to show that the action can be deformed to that of the B-twist
\cite{nm} of $N\!=\!4$ $d\!=\!4$ Yang-Mills theory.

\subsection{Super-BF and the B-Twist of $N\!=\!4$ $d\!=\!4$ Yang-Mills
Theory}

In \cite{ewdon} Witten pointed out that the topological Yang-Mills theory
he constructed (now known as Donaldson-Witten theory) could be regarded as
a twisted version of $N\!=\!2$ $d\!=\!4$ Yang-Mills theory. The `twist'
involves replacing one factor of the Lorentz group  $SU(2)_{L} \times 
SU(2)_{R}$ by its diagonal product with the internal R-symmetry group
$SU(2)_{I}$. Thus, the fermions, which originally transform as 
\be
\mbox{Fermions:} \;\;\;\; (2,1,2) \oplus (1,2,2)
\ee
under $SU(2)_{L}\times SU(2)_{R}\times SU(2)_{I}$, in the twisted theory
theory transform as, say,
\be
\mbox{Fermions} \ra (2,2) \oplus (1,1) \oplus (1,3)\;\;,
\ee
leading to a Grasssmann odd  vector, scalar and self-dual two-form
respectively. The emergence of the Grassmann odd scalar also reflects
the fact that that one of the supercharges has become a Lorentz scalar,
thus opening up the possibility to define the twisted theory on an
arbitrary curved manifold while preserving the scalar (topological)
supersymmetry.

This procedure was generalized to the $N\!=\!4$ $d\!=\!4$ Yang-Mills
theory by Yamron \cite{yamron}.  This is the dimensional reduction of
$N\!=\!1$ $d\!=\!10$ Yang-Mills theory to $d\!=\!4$.  As the $d\!=\!10$
spinor is Majorana-Weyl, there is no R-symmetry in $d\!=\!10$ and the
R-symmetry group of the $d\!=\!4$ theory arises exclusively from the
internal Lorentz group. Thus, decomposing the $N\!=\!1$ $d\!=\!10$
theory under $SO(10)\ra SO(4)\times SO(6)$, one finds that the
R-symmetry group is $SU(4)$.

Again, the twisting involves a choice of homomorphism from the Lorentz group
to the global symmetry group, and the result of the twisting is most
usefully and compactly expressed in terms of how the $(4)$ of $SU(4)$
decomposes under $SU(2)_{L}\times SU(2)_{R}$ \cite{vw}. 
There are three essentially
different possibilities giving rise to theories with scalar supercharges, 
arising from the branchings
\bea
(4) &\ra& (2,1) \oplus (1,2) \non
(4) &\ra& (1,2) \oplus (1,2) \non
(4) &\ra& (1,2) \oplus (1,1) \oplus (1,1)
\eea
respectively. As the fermions originally transform as 
\be
\mbox{Fermions:} \;\;\;\; (2,1,4) \oplus (1,2,\bar{4})
\ee
under $SU(2)_{L}\times SU(2)_{R}\times SU(4)$, descending from a $d\!=\!10$ 
Majorana-Weyl spinor with 16 real supercharges, in the twisted theory they
transform as, 
\bea
\mbox{Fermions} &\ra& 2(1,1) \oplus 2(2,2) \oplus (3,1) \oplus (1,3)\non
                &\ra& 2(1,1) \oplus 2(2,2) \oplus 2(1,3)\non
                &\ra& (1,1) \oplus (2,2) \oplus (1,3) \oplus 2(1,2) \oplus 
                      2(2,1)\;\;.
\label{4dtwist}
\eea
This exhibits the third theory as the half-twisted model of \cite{yamron},
Donaldson-Witten theory coupled to spinorial 
(hypermultiplet) matter. The second theory, 
with a doubling of the Donaldson-Witten field content and $N_{T}\!=\!2$, is
the A-model, i.e.\ 
the Euler character theory of instantons discussed in section 2 and 
constructed in this way by Yamron \cite{yamron} and Vafa-Witten \cite{vw}.

The first model, the B-model, is the one of interest here. Following
\cite{yamron,vw}, the B-model was constructed by Marcus 
\cite{nm}. Its scalar field content follows from the observation that
in the untwisted theory the scalars, as internal components of the $d\!=\!10$
connection, transform in the $(6)$ of $SU(4)\sim SO(6)$ which is the
antisymmetric product of the $(4)$. Thus,
\be
\mbox{Scalars} \ra \wedge((2,1)\oplus(1,2))=2(1,1) \oplus (2,2)
\ee
and the bosonic field content consists of a connection, two scalars, and a
one-form. 

We observe that this is precisely the (reduced) field content of $d\!=\!4$
super-BF theory discussed above (after elimination of the auxiliary 
fields $B$ and $u$), and in order to identify these two theories it 
remains to exhibit the $N_{T}\!=\!2$ symmetry in that model. 

Spelling out the action (\ref{sbf4act}) in detail one obtains
\bea
S_{q}^{(4),red}&=& 
Q\int\c F_{A} + d_{A}V*\c + Q\int d_{A}\fb*\p - d_{A}\eb *V\non
&=& \int BF_{A}+\c d_{A}\p + *\c d_{A}\pb + [\p,V]*\c + d_{A}V*B\non
&+& \int -(d_{A}\e *\p + d_{A}\eb *\pb)-d_{A}\fb*d_{A}\f + d_{A}u*V\non
&+& \int [\p,\fb]*\p - [\p,\eb]*V\non
\eea
(all commutators are to be understood in the graded sense). One observes
that there is almost a discrete symmetry exchanging $\p$ and $\pb$ and
$\e$ and $\eb$ with $(u,V,\c)\ra(-u,-V,*\c)$. To eliminate the `almost',
one can deform this action by adding to it
\be
\Delta S = Q\int -\trac{1}{2}[V,V]\c + \trac{i}{2}\c *B + [\pb,\fb]*V
\ee
(the factor of $i$ being required in Euclidean space so that the 
$B^{2}$-term damps the amplitude $\exp iS$). Shifting $B\ra B -id_{A}V$,
one finds
\bea
S &=& S_{q}^{(4),red}+\Delta S  \non
  &=& \int B(F_{A}-\trac{1}{2}[V,V]) +\trac{i}{2}(B*B+d_{A}V*d_{A}V)\non
  &+& \int d_{A}u*V - d_{A}\fb*d_{A}\f - \trac{i}{2}[\c,*\c]\f\non
  &+& \int (\c d_{A}\p +*\c d_{A}\pb) + 
      ([\pb,\e]*V-[\p,\eb]*V) -(d_{A}\e *\p + d_{A}\eb *\pb)\non
  &+& \int([\p,V]*\c -[\pb,V]\c) -(\fb[\pb,*\pb]+\fb[\p,*\p]) - [V,\f]*[V,\fb]
\;\;.
\eea
This action is now manifestly invariant under
\be
(A,\p,\pb,\f,V,\fb,\e,\eb,u,\c,B) \ra (A,\pb,\p,\f,-V,\fb,\eb,\e,-u,*\c,B) 
\ee
Consequently, in addition to the symmetry $Q$ displayed in 
(\ref{qred},\ref{qred2}), the action has a symmetry
$\bar{Q}$ following from combining $Q$ with this discrete symmetry. 
Thus, e.g.\ the $\chi$-transformation $Q\chi = B+id_{A}V$ (taking into
account the shift of $B$) gives rise to the $\bar{Q}$-transformation
$\bar{Q}\chi = *(B-id_{A}V)$. The complete set of transformations is
\bea
QA=\p&&\Q A=\pb\non
Q\p=d_{A}\f&&\Q\p=-[V,\f]\non
Q\f=0&&\Q\f = 0\non
Q\fb=\e&&\Q\fb=\eb\non
Q\e=[\fb,\f]&&\Q\e = -u\non
QV=\pb&&\Q V = -\p\non
Q\pb=[V,\f]&&\Q\pb=d_{A}\f\non
Q\eb=u&&\Q\eb=[\fb,\f]\non
Qu=[\eb,\f]&&\Q u = -[\e,\f]\non
Q\c=B + id_{A}V && \Q = *(B-id_{A}V)\non
QB = [\c,\f]-i[\p,V]+ id_{A}\pb&& \Q B = [*\c,\f]+i[\pb,V] + id_{A}\p
\eea
The operators $Q$ and $\Q$ are both equivariantly nilpotent, i.e.\ they
square to gauge transformations generated by $\f$,
\be
Q^{2}=\Q^{2}={\cal L}_{\f}\;\;.
\ee
They also anticommute, on-shell on $\c$ and $B$ and off-shell on all the
other fields,
\bea
\{Q,\Q\}\c &=& 2i \frac{\d S}{\d \c}\non
\{Q,\Q\}B &=& -2i [\frac{\d S}{\d B},\f]\non
\{Q,\Q\} &=& 0 \;\;\;\;\;\;\mbox{otherwise}\;\;.
\eea
To make contact with the action given by Marcus in \cite{nm}, one
can proceed to integrate out the field $B$. Then the first line of the
action $S$ becomes $(i/2)$ times
\be
\int (F_{A}-\trac{1}{2}[V,V])* (F_{A}-\trac{1}{2}[V,V])+d_{A}V*d_{A}V
=\int F_{A+iV}*F_{A-iV}\;\;.
\ee
This is precisely the bosonic part of the B-model action involving the
complexified connections $A\pm iV$. To compare the other terms in the 
action, one notes that Marcus' fields, indicated by a subscript $M$ in
the following, are related to those used here by complex linear combinations.
With the dictionary
\bea
Q_{M}=Q+i\Q&& \Q_{M}=Q-i\Q\non
\p_{M}=\pb -i\p&& \c_{M}=-\trac{1}{2}[\c+i*\c]\non
\e_{M}=\eb-i\e&& P_{M}=2u\non
C_{M}=\trac{1}{2}\fb&&B_{M}=4\f
\eea
one finds that the action and supersymmetry transformations given here
match precisely those of \cite{nm} in the $\alpha\ra 0$ limit.  
To obtain the action corresponding to $\alpha \neq 0$ one needs 
to add a term proportional to $u*u \sim - Q\Q\e*\eb$.             

Thus we have shown that the topological 
B-twist of $N\!=\!4$ $d\!=\!4$ Yang-Mills
theory studied in \cite{nm} can be regarded as a deformation of $d\!=\!4$
super-BF theory. At this particular point in the deformation space, the
super-BF action exhibits an extended $N_{T}\!=\!2$ supersymmetry.
Conversely, if one is willing to sacrifice one of the supersymmetries,
one can deform Marcus' action, which localizes onto complexified flat 
connections, to a theory which is manifestly real and localizes onto
the moduli space of real flat connections (or, rather, taking into
account the $V$ zero modes, its tangent bundle). Note that this deformation,
although it changes the moduli space, does not effect the topology of the
moduli space one is localizing on. As such it is a legitimate, albeit
somewhat unusual, deformation in the context of topological field theory.

In particular, formally observables and correlation functions 
in the two models agree, the additional $\Q$-symmetry in one
of them just serving to pick out a particular representative of the
$Q$-cohomology class. However, we will not dwell on this issue as
it is not clear to us that these correlation functions are mathematically
meaningful objects (for a discussion of some of the issues at stake 
see section 5.4.3 of \cite{bbrt} and section 5 of \cite{nm}). It is
thus equally unclear if calculations in this model can be used as a test
of S-duality in the spirit of \cite{vw}.

\subsection{A Novel Topological Twist of $N\!=\!4$ $d\!=\!3$ Yang-Mills Theory}

The point of departure this time 
is the $N\!=\!4$ $d\!=\!3$ Yang-Mills theory, i.e.\
the dimensional reduction of the minimal $N\!=\!1$ $d\!=\!6$
or $N\!=\!2$ $d\!=\!4$ Yang-Mills theory to $d\!=\!3$. This theory has a global
$SU(2)_{E}\times SU(2)_{N}\times SU(2)_{R}$ invariance (using the notation
employed in \cite{sw3d}), where $SU(2)_{E}$ is the Euclidean group in
$d\!=\!3$, $SU(2)_{N}$ is the internal Euclidean group arising from the
decomposition $SO(6)\ra SU(2)_{E}\times SU(2)_{N}$, and $SU(2)_{R}$ is the
R-symmetry group of the six-dimensional theory (see \cite{n4d3} for other 
recent work on $N=4$ $d=3$ theories). 

The field content of the untwisted dimensionally reduced model is 
a gauge field $A$ and fermions and three scalars transforming as 
\bea
\mbox{Fermions:}&& (2,2,2)\non
\mbox{Scalars:}&& (1,3,1) 
\eea
of $SU(2)_{E}\times SU(2)_{N} \times SU(2)_{R}$. There are two obvious ways
of topologically twisting this model.      

Twisting the $d\!=\!3$ Lorentz group $SU(2)_{E}$ by $SU(2)_{R}$ leads to the 
field content of the $d\!=\!3$ super-BF (or Casson, or Euler character, or
super-IG) model \cite{ewtop,bbt,bbrt,btmq}, 
namely a gauge field and twisted fermions and scalars
transforming as
\bea
\mbox{Fermions} &\ra& (1,2)\oplus (3,2)\non
\mbox{Scalars} &\ra& (1,3)\;\;,
\eea
the second slot indicating the transformation behaviour under the
remaining unbroken $SU(2)_{N}$.
This is not surprising as this is just the dimensional reduction of
Donaldson-Witten theory (the same twist being used in both cases). 
Interpreting the $SU(2)_{N}$ as a ghost-number symmetry, one can
read off from the above that the model has an $N_{T}\!=\!2$ topological
supersymmetry, as indeed we knew, the two charges $Q$ and $\Q$ transforming 
as a doublet of $SU(2)_{N}$. 

The second twist, by $SU(2)_{N}$, is intrinsically three-dimensional
and leads to the field content
\bea
\mbox{Fermions}&\ra& (1,2)\oplus (3,2) \non
\mbox{Scalars}&\ra& (3,1)\;\;. 
\eea
Consequently, as one has twisted by the internal Lorentz group and the
scalars are the internal component of the $d\!=\!6$ gauge field, in this
twisted theory the three scalars combine to form a $d\!=\!3$ one-form
we will call $V$. Thus this theory has the somewhat unusal property of
possessing no bosonic scalars. The other fields are the gauge field $A$, 
Grassmann odd scalars $\e$ and $\eb$, and Grassmann odd one-forms $\p$ 
and $\c$. Note that, once again, this theory will have an $N_{T}\!=\!2$
supersymmetry, the two supercharges $Q$ and $\Q$ now transforming as 
a doublet of the R-symmetry group $SU(2)_{R}$. 

As this twisted model differs from the first one by an exchange of
$SU(2)_{R}$ and $SU(2)_{N}$, it is tempting to speculate that it can
be regarded as providing a mirror description of the Casson invariant 
in the spirit of \cite{n4d3}. However, at present we have no solid 
evidence in favour of this.

One rather direct way of obtaining the action and supersymmetries of this 
model is to dimensionally reduce and twist the $N\!=\!1$ $d\!=\!6$ Euclidean 
Yang-Mills theory. The details of this procedure are straightforward but
not particularly interesting, so we will only fill in some of the steps.

The Euclidean $N\!=\!1$ $d\!=\!6$ action is 
\be
L_{E} = -\pdr\G^{M}D_{M}\pl -\trac{1}{4}F_{MN}F^{MN}
\ee
where the fermions are chiral spinors in the $(4)$ and $(\bar{4})$ of 
$Spin(6)=SU(4)$ and (in Euclidean space) the right-handed spinor $\pdr$ 
is taken to be independent of $\pl$. The $\G^{M}$ are Euclidean gamma 
matrices, 
\be
\{\G^{M},\G^{N}\} = 2 \d^{MN}\;\;.
\ee
This action has the two independent supersymmetries
\bea
\d\pl &=& \trac{1}{2}\G^{M}\G^{N}F_{MN}\eps_{L}\non
\d\pdr &=&-\trac{1}{2}\eps_{R}^{\dag}\G^{M}\G^{N}F_{MN}\non
\d A_{M} &=& - (\eps^{\dag}_{R}\G_{M}\pl -  \pdr\G_{M}\eps_{L})\;\;.
\eea
We will use a representation of the gamma
matrices in $d\!=\!6$ which is well adapted to a $6=3+3$ decomposition.
The basic building blocks are the Pauli matrices
\be
\sigma_{1} = \mat{0}{1}{1}{0}\;\;,\;\;\;\;
\sigma_{2} = \mat{0}{-i}{i}{0}\;\;,\;\;\;\;
\sigma_{3} = \mat{1}{0}{0}{-1}\;\;,
\ee
satisfying
\bea
\sig_{k}\sig_{l} = i \eps_{klm}\sig_{m} + \d_{kl} \II\;\;,\non
\sk^{2} = \II\;\;,\;\;\;\;\;\;\sk^{\dag}=\sk\;\;.
\eea
In terms of these, and the $(2\times 2)$ identity matrix $\II$, 
one choice for the $(8\times 8)$ gamma matrices is
\bea
\Gamma_{k} &=& \sigma_{1}\ot\II\ot\sigma_{k}\non
\Gamma_{k+3}\equiv\Gamma_{a} &=& \sig_{2}\ot\sa\ot\II\non
\Gamma_{7}&=& \sig_{3}\ot\II\ot\II\;\;. 
\eea
where the convention for tensor products is such that e.g.\
$\sig_{1}\ot\II$ denotes the $(4\times 4)$ matrix
\be
\sig_{1}\ot\II = \mat{0_{2}}{\II}{\II}{0_{2}}\;\;.
\ee
Hence a $d\!=\!6$ Weyl spinor is of the form $\Psi_{L}=(\psi,0)^{t}$, where
$\psi$ is a four-component complex spinor and an independent
right-handed conjugate spinor is of the form $\pdr=(0,\chi)$.

The anti-hermitian generators of $Spin(6)$ are thus
\bea
\trac{1}{2}[\G_{k},\G_{l}] &=&\II\ot\II\ot i\eps_{klm}\sig_{m}\non
\trac{1}{2}[\G_{a},\G_{b}] &=&\II\ot\eps_{abc}\sig_{c}\ot\II \non
\trac{1}{2}[\G_{k},\G_{a}] &=&i\sig_{3}\ot\sig_{a}\ot\sk\;\;.
\eea

It follows from the above that under $SU(2)_{N}\times SU(2)_{E}
\ss Spin(6)$, the chiral spinor $\p$ 
transforms as a $(2,2)$, with $SU(2)_{E}$, generated by $[\G_{k},\G_{l}]/2$,
acting
diagonally on the two 3d spinors (i.e.\ the first two and last two
components respectively), while they transform as a doublet under
$SU(2)_{N}$. In other words, the action of $SU(2)_{N}\times SU(2)_{E}$ is
generated by $\sig_{a}\ot\sig_{k}$, and in terms of components $\psi_{A'B}$
one has
\be
\psi_{A'B}\ra\sig_{aA'}^{\;\;C'}\sig_{kB}^{\;\;D}\psi_{C'D}\;\;.
\ee
We will now take all fields to be independent of $x^{4},x^{5},x^{6}$.
After twisting $SU(2)_{E}$ by $SU(2)_{N}$, the internal components
$A_{c}$ of the connection $A$ transform as a covector $V_{k}$ while 
the fermions transform as a scalar and a one-form,
\bea
\psi_{AB}&=&\eps_{AB}\e + \psi_{k}\sig^{k}_{AB}\non
\chi^{AB}&=&\eps^{AB}\eb + \chi_{k}\sig^{k\;AB}\;\;,
\eea
where $\eps_{12}=\eps^{12}=1$ and $\sig^{k}_{AB} = \sig^{k}_{A}{}^{C}
\eps_{CB}$ is symmetric.

The bosonic part of the Lagrangian becomes   
\bea 
L_{b} &=& -\trac{1}{4}F_{MN}F^{MN}\non
      &\ra& -\trac{1}{4}(F_{ij}-[V_{i},V_{j}])^{2} -\trac{1}{4}
(D_{i}V_{j}-D_{j}V_{i})^{2} -\trac{1}{2}(D_{i}V^{i})^{2}\non
&=& -\trac{1}{4}F_{ij}(A+iV)F^{ij}(A-iV)-\trac{1}{2}(D_{i}V^{i})^{2}\;\;,
\eea
and from the Dirac action one obtains
\bea
L_{f}&=& -\c^{AB}\d_{A}^{C}\sig_{kB}{}^{D}D_{k}\p_{CD}
         -i\c^{AB}\sig_{aA}{}^{C}\d_{B}^{D}D_{a}\p_{CD}\non
     &\ra& 2\eb D_{k}(A-iV)\p_{k}-2\c_{k}D_{k}(A-iV)\e \non
     && +2i\eps_{lkm}\c_{l} D_{k}(A+iV)\p_{m}\;\;.
\eea
Hence, putting the two together 
one finds that the action of the twisted 3d model is
\bea
S&=&-\trac{1}{2}\int F_{A+iV}*F_{A-iV} + d_{A}*V*d_{A}*V\non
 &&-2\int \eb d_{A-iV}*\p-\e d_{A-iV}*\c - i\c d_{A+iV}\p\;\;.
\eea
To obtain the topological symmetries of this action, one needs 
to work out the scalar components
$\eps_{L,AB} = \eps_{L}\eps_{AB}$ and $\eps_{R}^{\dag}{}^{AB}
=\bar{\eps}\eps^{AB}$ of the supersymmetry transformations. 
Denoting the corresponding BRST operators
by $\bar{Q}$ and $Q$ respectively, one finds
\bea
Q(A+iV) = 4\p && \bar{Q}(A+iV)=-4\c \non
Q(A-iV) = 0 && \bar{Q}(A-iV)=0 \non
Q\p=0 && \bar{Q}\psi = -i*F_{A-iV}\non
Q\c = -i*F_{A-iV} && \bar{Q}\c = 0\non
Q\e = 0&& \bar{Q}\e = -i*d_{A}*V\non
Q\eb = -i*d_{A}*V&& \bar{Q}\eb = 0 \;\;.
\eea
Note that the action has the discrete symmetry
\be
(\p,\c,\e,\eb)\ra (-\c,\p,-\eb,\e)
\ee
mapping $Q$ to $\Q$, so that $Q$-invariance is equivalent to $\Q$
invariance.  This symmetry can be interpreted as
the action of the Weyl subgroup of the R-symmetry group $SU(2)_{R}$.
This can be seen by introducing the $SU(2)$ doublets 
$i\EE^{A}=(\eb,\e)$ and $\PP^{A}=(\c,\p)$ in terms of which the
fermionic Lagrangian can be written as 
\be
i\epsilon_{AB}(\EE^{A}d_{A-iV}*\PP^{B}-\frac{1}{2}\PP^{A}d_{A+iV}\PP^{B})\;\;.
\ee 
It is straightforward to see that indeed  
\be
QS=\bar{Q}S = 0\;\;.
\ee
This can be made manifest by introducing some auxiliary fields which 
have the added virtue of making $Q$ and $\bar{Q}$ nilpotent and anticommuting
off-shell (so far they do so only modulo the $\e$ and $\eb$ equations of
motion). Thus, we introduce two auxiliary one-forms $B$ and $\Bb$, and
a scalar $u$ with transformation rules (designed to preserve the above
discrete invariance supplemented by $(B,\Bb,u)\ra(B,\Bb,u)$)
\bea
Q\c=B&&\Q\p=B\non
QB=0&&\Q B=0\non
Q\eb=u&&\Q\e=u\non
Qu=0&&\Q u =0\non
Q\Bb=-d_{A-iV}\e && \Q\Bb=d_{A-iV}\eb\;\;.
\eea
Then one has 
\be
Q^{2}=\Q^{2}=\{Q,\Q\}=0
\ee
off-shell. The action can now be written as a sum of a topological (BF like)
term and a term which is actually $Q\Q$-exact,
\be
S= 2i\int \Bb F_{A-iV} +\trac{i}{16}Q\Q S_{CS}(A+iV) -
Q\Q\int iV*\Bb+\trac{1}{2}\e *\eb\;\;.
\ee
Here $S_{CS}$ is the Chern-Simons action (with normalization $\int
AdA+\ldots$), and the first term is $Q$ and $\Q$ invariant by the Bianchi
idenity. Integrating out $B,\Bb$ and $u$ reproduces the action and
transformation rules given above.

While we will leave a more detailed investigation of the localization
properties and correlation functions in this model to future investigations,
we do want to point out two rather striking features of this model
which set it apart both from the B-model, discussed above, to which it 
bears a superficial resemblance, and from other known 
cohomological gauge theories. 

One of the unusual features is that the
supersymmetry is not equivariant in any sense (and indeed it hardly
could be in the absence of the usual scalar ghost for ghost $\f$), but
rather strictly nilpotent even prior to the introduction of gauge ghosts.

The second remarkable property is that $A-iV$ is both $Q$ and
$\bar{Q}$ invariant. Hence any gauge invariant functional of $A-iV$,
constrained by $F_{A-iV}=0$, is a candidate observable. In particular, here
we have a cohomological theory in which bosonic Wilson loops appear to be
good observables. 

Finally, note that there is no net ghost number violation, as it should 
be since the unbroken R-symmetry group $SU(2)_{R}$ is simple. Hence the 
partition function and correlation functions of Wilson loops of 
$A-iV$ are potentially non-vanishing.

\section{D-brane Instantons and Topological Gauge Theories in $d\!=\!2,3$}

In this section, we will discuss some other classes of low-dimensional
topological gauge theories, arising from the dimensional reduction of
$N\!=\!1$ $d\!=\!10$ or 
$N\!=\!4$ $d\!=\!4$ Yang-Mills theory to $d\!=\!3$ and $d\!=\!2$. As has
been explained e.g.\ in \cite{ewpb}, the 
dimensional reduction of $N\!=\!1$ $d\!=\!10$
super-Maxwell theory to $(p+1)$ dimensions describes the effective
low-energy world-volume dynamics (i.e\ the collective coordinates) of
flat Dirichlet $p$-branes (with enhanced $U(n)$ gauge symmetry for 
coincident D-branes). 

Of course, the importance of the study 
of D-branes \cite{jp1}
for a deeper understanding of string theory need not be   
stressed and we just refer to two recent extensive reviews on D-branes
\cite{jptasi} and string dualities \cite{jsictp} and the references
therein for further information. What we will focus on in the following
is a beautiful observation due to Bershadsky, Sadov and Vafa \cite{bsv}
that topologically twisted versions of these supersymmetric world-volume
theories appear completely naturally in the study of curved D-branes
and in particular for D-brane instantons wrapping around supersymmetric
cycles \cite{bbs} of the compactifying space. In particular, via an  
argument that we will recall below, they showed that all the three
different twistings of $N\!=\!4$ $d\!=\!4$ Yang-Mills theory exhibited in
(\ref{4dtwist}) appear in this way as effective world-volume theories, 
namely for special Lagrangian submanifolds of Calabi-Yau four-folds
(the B-model \cite{nm}), for coassociative submanifolds of $G_{2}$-holonomy
seven-manifolds (the A-model of \cite{btmq} and Vafa-Witten
\cite{vw}),  and for Cayley-submanifolds of Spin(7) eight-manifolds
(the half-twisted model). Furthermore, the partial twist along a 
two-dimensional surface considered in \cite{bjsv} appears for three-branes
wrapping a two-cycle of a K3.  

With this in mind, we will in the following analyse the topological 
twistings of the $d\!=\!3$ theories. It turns out that there are only
two (partial) topological twists (provided that one excludes theories
involving higher spin fields) and that their field content and
supersymmetries are precisely what one expects for the two known classes
of supersymmetric three-cycles, namely special Lagrangian submanifolds
of Calabi-Yau three-folds and associative submanifolds of $G_{2}$-holonomy
manifolds. We regard it as rather pleasing (and intriguing) that also in 
this case all the topological twists can be naturally realized in this 
manner.

We will also briefly describe the topological gauge theories associated
to Dirichlet one-brane (D-string) instantons wrapping holomorphic
curves in K3s and Calabi-Yau three-folds.

Before proceeding to these lower dimensions, we would like to point out
that twisted models can also be constructed under certain conditions
in $d=5$
and $d=6$. As emntioned in the introduction, these appear to be related to
considerations in \cite{lmns} and \cite{dvv} respectively. It should be
interesting to study this further.

\subsection{Topological Twists of $N\!=\!8$ $d\!=\!3$ Yang-Mills Theory}

In this section, the theory of interest is once again the 
dimensional reductions of $N\!=\!1$ $d\!=\!10$ YM theory, 
this time to $d\!=\!3$.

Thus the global symmetry group is
\be
H=SU(2)_{E}\times Spin(7)\;\;,
\ee
under which the gauge field $A$, spinors and scalars transform as
\bea
\mbox{Gauge field:}&& (3,1)\non
\mbox{Fermions:}&& (2,8)\non
\mbox{Scalars:}&& (1,7)\;\;,
\eea
where $(8)$ is the spinor representation of $Spin(7)$ and $(7)$ is the
vector representation. 

Twists in $d\!=\!3$ involve decomposing the $(8)$ of $Spin(7)$ under $SU(2)$.
Clearly {\em a priori} there are many possibilities. However, the 
requirement that the spinor, which is a $(2,8)$, turn into something sensible
in the twisted theory severely restricts the number of viable options.

First of all, in order that the twisted theory contain at least one
scalar supercharge, a $(2)$ of $SU(2)$ must occur in the decomposition 
of $(8)$, 
\be
(8) \ra (2) \oplus_{i}R_{i}\;\;.
\ee
Furthermore, among the $R_{i}$ no representations of spin $\geq 1$ should
appear, because otherwise spin $>1$ fermionic fields will appear in the
twisted action. E.g.\ if one of the $R_{i}$ were a $(3)$ of spin one, from
$(2) \ot (3) = (2) \oplus (4)$ one would obtain a spin $3/2$ field in the 
$(4)$
of $SU(2)$. Hence, the second requirement is 
\be
\mbox{dim}\;R_{i}\leq 2\;\;.
\ee
Finally, for a full (as opposed to partial) topological twist,
only half-integral spins should appear among the $R_{i}$ so that
the $(2) \ot R_{i}$ are all tensorial. 

A systematic search, using one's favourite reference on branchings
(see e.g.\ \cite{branchings}), reveals that there are essentially
only two possibilities satisfying the first two desiderata, namely
$(8)\ra 4(2)$ and $(8)\ra 2(2)\oplus 4(1)$, the former 
also satisfying the third and leading to a full topological
twist. Both of them are most transparently described in terms of the 
branching
\be
Spin(7) \ra SU(2) \times SU(2) \times SU(2)\;\;,
\label{7222}
\ee
which will make the maximal residual global symmetry group $SU(2) \times
SU(2)$ manifest. Under (\ref{7222}), the $(8)$ and $(7)$ of $Spin(7)$ 
decompose as
\bea
(8) &\ra& (2,1,2) \oplus (1,2,2)\non
(7) &\ra& (2,2,1) \oplus (1,1,3)\;\;.
\eea
Then, taking e.g.\ the diagonal of $SU(2)_{E}$ with the right-most 
$SU(2)$-factor one finds that in the twisted theory the spinors and scalars 
transform as
\bea
(2,8) &\ra& (1,2,1)\oplus (1,1,2) \oplus (3,2,1) \oplus (3,1,2)\non
(1,7) &\ra& (1,2,2) \oplus (3,1,1)
\label{d3n4}
\eea
under $SU(2)_{E}\times SU(2)\times SU(2)$. Therefore this twisted theory
has $N_{T}\!=\!4$, the four scalar supercharges transforming as two $SU(2)$
doublets. The field content consists of the $d\!=\!3$
gauge field $A$ plus four
Grassmann odd scalars $\e^{A}$ and $\eb^{A'}$, four Grassmann odd vectors
$\p^{A}$ and $\pb^{A'}$ as well as four bosonic scalars $\f^{AA'}$ and one
bosonic one-form $V$ (which is an $SO(4)$-singlet). 

It is easy to see that this toplogical theory is precisely the dimensional
reduction of either of the two $d\!=\!4$ $N_{T}\!=\!2$ theories. That both of
these reduce to equivalent $d\!=\!3$ theories explains why there are only two
topological twists of the $d\!=\!3$ theory whereas one might have expected at
least three - arising from the dimensional reduction of the three topological
twists of $N\!=\!4$ $d\!=\!4$ Yang-Mills theory. 

The other topological twist is obtained by taking the diagonal of $SU(2)_{E}$
with the first of the three other $SU(2)$-factors. In this case one finds
\bea
(2,8) &\ra& (1,1,2) \oplus (3,1,2) \oplus (2,2,2)\non
(1,7) &\ra& (2,2,1) \oplus (1,1,3)\;\;.
\label{d3n2}
\eea
Thus this is a partially twisted $N_{T}\!=\!2$ theory, and indeed precisely
the dimensional reduction of the half-twisted 
$N_{T}\!=\!1$ $d\!=\!4$ theory.
Its field content consists of the $d\!=\!3$ gauge field $A$ plus two Grassmann
odd scalars $\e^{A'}$, two Grassmann odd vectors $\p^{A'}$, four spinors
$\lambda^{AA'}$, three scalars $\f^{i'}$ (transforming as a vector under
the second internal $SU(2)$) and two scalar spinors $\beta^{A}$. 

This $N_{T}\!=\!2$ theory can for instance also be described in terms of the
branching 
\be
Spin(7) \ra G_{2} \ra SU(2) \times SU(2)\;\;.
\ee
In that description, however, only the diagonal subgroup of the global
symmetry group $SU(2)\times SU(2)$ is manifest. 

The action of this theory can be described as the coupling of the 
$N_{T}\!=\!2$
$d\!=\!3$ Euler character (super-BF, Casson, \ldots) theory to a 
hypermultiplet.
Indeed the fields can be put into $N_{T}=2$ superfields of 
section 3 augmented with $N_{T}=2$ spinor superfields, thus making the 
topological invariance and symmetry manifest. 
We have not been able to find a superfield formulation for the $N_{T}=4$ 
theory based on the superfields of section 3.6. However, a natural 
expectation is that for this theory (and indeed for all the 
topological 
theories that come from twisting reductions of the $N=1$ $d=10$ theory) 
one simply needs to twist the $10$ 
dimensional superfields after dimensional reduction to obtain the 
topological superfields with manifest, but perhaps on-shell, 
supersymmetry.

Finally, we want to point out that one can construct precisely
two other theories
with $N_{T}>0$ by twisting, both of which however involve fields transforming
in the $(4)$ of the twisted Lorentz group, i.e.\ spin-3/2 Rarita-Schwinger
fields. An $N_{T}\!=\!2$ theory of this kind can, for instance, be obtained
by twisting once more the above $N_{T}\!=\!4$ theory with one of the internal
$SU(2)$s. The $N_{T}\!=\!1$ theory follows e.g.\ from the chain of
branchings
\be
\begin{array}{lllllll}
Spin(7)& \ra& SU(4)& \ra&  SU(3)& \ra&  SU(2)\\
(8) & \ra & (4) \oplus (\bar{4}) & \ra & (3) \oplus (\bar{3}) \oplus 2(1) &
\ra & (3) \oplus (2) \oplus 3(1)\\
(7) & \ra & (6)+(1)& \ra & (3)\oplus (\bar{3}) \oplus (1) & \ra & (3)
\oplus (2) \oplus 2(1)
\end{array}
\ee
under which the spinors and scalars can be twisted to
\bea
(2,8) &\ra& (1) \oplus 4(2) \oplus (3) \oplus (4)\non
(1,7) &\ra& 2(1) \oplus (2) \oplus (3)\;\;.
\eea
This observation may be of interest in its own right as consistent 
interacting higher spin theories
are hard to come by even in flat space, while the twisting should make
no difference there and thus map the consistent untwisted theory to another
consistent theory. On the other hand it is 
certainly not guaranteed by the twisting procedure alone that
these theories will make sense on a curved manifold.

\subsection{Realizations as World-Volume Theories of 2-Brane Instantons}

We now want to show that the two topological theories found above are
naturally realised as world-volume theories of Dirichlet 2-brane
instantons in type IIA string theory. For other work on Dirichlet 
$p$-brane instantons see e.g.\ \cite{bbs,mol,ovss,ooy,bbmooy}.

The bosonic world volume fields of a (flat) Dirichlet
p-brane are a world volume vector field $A$, arising from the boundary
conformal field theory or the Chan-Paton factors of the open string
theory, and $10-(p+1) = 9-p$ scalar fields, the collective
coordinates or Goldstone modes for the broken translation invariance.
Now one can consider the situation where a curved Euclidean D-brane 
`wraps' around a non-trivial cycle $S$ of the compactifying manifold $M$
(e.g.\ a Calabi-Yau manifold). For such a configuration to preserve some
fraction of the supersymmetries left unbroken by the compactification,
i.e.\ for the cycle to be supersymmetric \cite{bbs}, the cycle needs to
satisfy some rather stringent conditions identifying it 
\cite{bbs,bsv,ooy,bbmooy} as a calibrated submanifold \cite{hl,mclean}.

For later reference, let us collect here the relevant information
concerning special-holonomy Ricci-flat manifolds.       
In the following table we have indicated  
the dimension $m$ of the manifold, the holonomy group, the name
usually given to such a manifold, the 
number of covariantly constant spinors (corresponding to the number $N$
of supersymmetries in the compactified theory) and the fraction of 
supersymmetries thus preserved by a particular compactification.
\be
\begin{array}{cllcc}
m=4 & SU(2) & \mbox{K3} & N\!=\!1 & 1/2 \\
m=6 & SU(3) & \mbox{Calabi-Yau} & N\!=\!1 & 1/4 \\
m=7 & G_{2}& \mbox{Joyce} & N\!=\!1 & 1/8 \\
m=8 & SU(4) & \mbox{Calabi-Yau} & N\!=\!2 & 1/8 \\
m=8 & Spin(7)& \mbox{Joyce} & N\!=\!1 & 1/16 
\end{array}
\ee
String compactifications on the first two types of manifolds are
of course well known. Calabi-Yau four-folds appear in compactifications
of F-theory \cite{cvf,cy4} and M-theory \cite{ewsp}, and some aspects of 
compactifications of string 
theory and M-theory on Joyce manifolds have been studied in 
\cite{pt,bsa,bbmooy}. 

Let us now recall the argument of \cite{bsv}. 
In general, even though a cycle may be supersymmetric it may not
possess any covariantly constant spinors. Thus the supersymmetry 
of the world-volume theory cannot in general be realized in the standard 
form but will have to involve some twisted definition of the supercharges
in order to give a meaning to the world-volume theory. 

This comes about as follows: As the scalars are associated with translations
of the D-brane, there should be only $10 -\mbox{dim}\,M\leq 9-p$ true scalars,
while the remaining translational modes should organize themselves into a
section of the normal bundle $N_S$ to $S$ in $M$. Thus these scalars are
`twisted' if the normal bundle is non-trivial, and so are then their
superpartners, the fermions. The number $N_{T}$ of scalar supercharges of 
the theory is obtained by matching $N_{T}/16$ with the fraction
appearing in the last column of the above table. 
Thus, for a given supersymmetric cycle, 
knowledge of its normal bundle in, and the number of covariantly constant 
spinors on, the ambient manifold determine the bosonic field content and
number of scalar supercharges of its $d=(p+1)$ dimensional
world-volume dynamics. Conversely,
given a (partially) topological gauge theory one can check if there are
supersymmetric cycles with the requisite properties. 

A slight refinement of these arguments also leads to information about
some global symmetries a topological gauge theory arising in this way 
should possess. Namely, thinking of it as arising from compactification
on $M$, there should be a global invariance under the rotation group
$SO(10-\mbox{dim}\,M)$ (or $SO(9-\mbox{dim}\,M,1)$) of the uncompactified
dimensions. In particular, the true scalars should organize themselves
into a vector in the fundamental representation of this group, with the
other bosonic fields being singlets. This gives some {\em a priori} 
conditions 
on which branchings of the R-symmetry group $SO(9-p)$ can lead to topological
theories associated with some manifold $M$ - those that proceed via the
branching 
\be
SO(9-p) \ra SO(10-\mbox{dim}\,M) \times SO(\mbox{dim}\,M -p-1)
\label{geob}
\ee
and subsequent twisting by the second factor (so that the first factor
is preserved and only the normal directions to $S$ in $M$ are affected
by the twisting). This is indeed what we will find below.

As an example, consider the B-model of section 4.2. Its bosonic field content
consists of the world-volume gauge field as well as of two true scalars and
one four-vector $V$. The fact that $d=(p+1)=4$ and that there are two scalars 
together imply that one is looking for a
four-cycle $S$ in an eight-dimensional manifold $M$. $N_{T}\!=\!2$ indicates 
that one is looking for an eight-manifold preserving $2/16 = 1/8$ of the 
supersymmetry. Thus $M$ is a Calabi-Yau four-fold and $S$ is a special
Lagrangian submanifold, i.e.\ a submanifold for which the real part of the
holomorphic four-form restricts to the volume form.
A consistency check is provided by the fact that
the normal bundle $N_S$ can in that case indeed be identified with the
cotangent bundle \cite{mclean}, accounting for the fact that the remaining 
four scalars make up  a one-form $V$ on $S$. Finally, we note that this
theory has a global $U(1)$ ghost-number symmetry, and that - in accordance
with the observation in the previous paragraph - this global symmetry can be
identified with the rotation group $SO(2)$ in the two uncompactified 
dimensions, the two true scalars having $U(1)$-charges $(2,-2)$ and hence
comprising a two-dimensional vector.
In a similar fashion, the 
other two topological twists of 
$N\!=\!4$ $d\!=\!4$ Yang-Mills can be identified 
with world-volume theories of D-brane instantons \cite{bsv} leading to the
result recalled at the beginning of this section. 

With this in mind, let us now turn to the two three-dimensional topological
theories obtained above. In the $N_{T}\!=\!4$ theory, instead of the seven true
scalars of the untwisted theory we have only four true scalars and one
one-form $V$. Thus we are looking for a six-manifold $M$ which preserves
$4/16=1/4$ of the supersymmetries, i.e.\ a Calabi-Yau three-fold. As in
the $d\!=\!4$ case above, the normal bundle of a special Lagrangian submanifold
$S$ in $M$ can be identified with the contangent bundle $T_{S}^{*}$, in
agreement with the appearance of $V$ as the remaining bosonic field.
Note that indeed, as anticipated above, the internal $SU(2) \times SU(2)$ 
can be thought of as the
$SO(4)$ of the uncompactified dimensions, the corresponding four scalars
transforming as a vector of $SO(4)$ while $V$ is an $SO(4)$-singlet, 

In the $N_{T}\!=\!2$ theory, on the other hand, we have three true scalars
transforming as a vector of one internal $SU(2)$ and two spinors $\beta^{A}$
transforming as a doublet of the other. By the now familiar
argument we expect to be dealing with a seven-manifold possessing one
covariantly constant spinor (i.e.\ preserving $1/8$ of the supersymmetries).
Thus $M$ is a Joyce \cite{joyce} seven-manifold of $G_{2}$-holonomy.

It is known \cite{mclean} that an associative $d\!=\!3$ submanifold $S$ of
a $G_{2}$ 7-manifold has a normal bundle $N\!=\!\SS\ot V$, where $\SS$ 
is the spinor bundle of $S$ and $V$ is a rank two $SU(2)$-bundle. But
this fits in perfectly with the fact that the twisted scalars $\beta^{A}$
are an $SU(2)$-doublet of spinors on $S$. Note also that once again the 
three true scalars organize themselves into a vector of the orthogonal group
$SO(3)$ of the uncompactified dimensions. 

Thus we have sucecssfully identified the two topological twists of
$N\!=\!4$ $d\!=\!3$ Yang-Mills theory with the world-volume theories of
Dirichlet two-brane instantons wrapped around the two known classes
of supersymmetric three-cycles.

That these identifications are correct can also be deduced from the 
dimensional reduction of the results of \cite{bsv} using the 
local models (coordinate representations) 
of calibrated manifolds given e.g.\ by McLean \cite{mclean}.

Let us start with the half-twisted $N_{T}\!=\!2$ model.
Recall that, according to 
\cite{bsv}, the $d\!=\!4$ half-twisted model corresponds to a so-called
Cayley submanifold of a $Spin(7)$ 8-manifold, characterized by a
Cayley 4-form. It is easy to see that integration of this form 
over a toroidal fibre produces the three-form characterizing
associative submanifolds of $G_{2}$-holonomy 7-manifolds, in 
agreement with the above identification.

Concretely this means the following: Let the Cayley 4-form in local
coordinates (or globally on the frame bundle) be given by ($[0123]$ 
stands for $dx^{0}dx^{1}dx^{2}dx^{3}$ etc.\ and the product is the
wedge product) 
\bea
\Omega_{Cayley} &=& [0123] + ([01]-[23])([45]-[67]) + ([02]-[13])([46]+[57])
\non && + ([03]-[12])([47]-[56]) + [4567]
\eea
(see (\cite[eq.\ (6.1)]{mclean}). 
Now integrate over the fibre $x^{0}$ to obtain
\be
\pi_{*}\Omega_{Cayley} = [123] + [1]([45]-[67])+[2]([46]+[57])
+[3]([47]-[56])\;\;.
\ee
According to \cite[eq.\ (5.1)]{mclean} this is precisely the three-form
$\Omega_{ass}$ characterizing associative 3-manifolds of $G_{2}$-manifolds,
\be
\pi_{*}\Omega_{Cayley} = \Omega_{ass}\;\;.
\ee

For the $N_{T}\!=\!4$ model 
there are two different 4-dimensional origins, as both $N_{T}\!=\!2$ $d\!=\!4$ 
theories reduce to it in $d\!=\!3$. On the one hand,
according to \cite{bsv}, the Euler character theory of \cite{btmq} and 
\cite{vw}, i.e.\ the A-model, corresponds to coassociative 
submanifolds of $G_{2}$-manifolds, 
characterized by the
Hodge dual 4-form $\Omega_{coass}= *\Omega_{ass}$ which, by a relabelling, 
we will write as (cf.\ \cite[eq. (4.4)]{mclean}) 
\be
\Omega_{coass} = [0123]-[56]([01]-[23]) +[46]([02]+[13])-[45]([03]-[12])\;\;.
\ee
Integrating this over $x^{0}$, one obtains 
\be
\pi_{*}\Omega_{coass}=[123]-[156]+[246]-[345]\;\;.
\ee
This ought to be compared with the (real part of the) holomorphic volume form 
of a Calabi-Yau 3-fold which characterizes special Lagrangian submanifolds.
The local model for this is the following. Let $z^{k}=x^{k}+ix^{k+3}$ be
local complex coordinates. Then
\bea
\Omega_{CY}  &=& dz^{1}\wedge dz^{2}\wedge dz^{3}\\
             &=& ([123]-[453]-[156]-[426]) + i([423]+[513]+[612]-[456])
\;\;,
\nonumber
\eea
so that indeed
\be 
\pi_{*}\Omega_{coass} = \mbox{Re}\Omega_{CY}\;\;,
\ee
as expected.

On the other hand, according to \cite{bsv}, the B-model is realized on
special Lagrangian submanifolds of Calabi-Yau 4-folds. This situation is
slightly different as now the dimension of the ambient space changes by
two in comparison with special Lagrangian submanifolds of Calabi-Yau
3-folds featuring in the dimensionally reduced theory. 
But if one assumes that the 4-fold is locally of the form
$CY_{4}=CY_{3}\times T^{2}$ (i.e.\ it is an elliptic fibration), then
clearly special Lagrangian submanifolds of $CY_{4}$ 
wrapping around one of the circles reduce (upon double dimensional reduction) 
to special Lagrangian submanifolds of $CY_{3}$, 
showing again the consistency of the dimensional reduction procedure 
with the geometrical interpretation in terms of calibrated submanifolds.

\subsection{Topological Gauge Theories on Holomorphic Curves}

In a similar spirit one can now discuss the theories associated with
Dirichlet one-brane (or D-string) instantons. Ignoring the world-volume
gauge field (which has no local dynamics) one would, by $SL(2,\ZZ)$ duality
of the type IIB string, expect D-string instantons to correspond to
standard world-sheet instantons. And indeed the only known supersymmetric
two-cycles are holomorphic curves in Calabi-Yau $n$-folds. In the following
we will focus on the two cases $n=2$ and $n=3$ and briefly
illustrate how the various facets of the corresponding world-volume theories
we have encountered for $d=3$ above fall into place here.

The starting point is again the dimensional reduction of $N\!=\!1$ $\!d=\!10$
Yang-Mills, this time to $d\!=\!2$. The resulting theory has $N\!=\!8$ 
supersymmetry,
and the R-symmetry group is now $SO(8)$ or $Spin(8)$.
The field content consists of a $d=2$ gauge field $A$ and the fermions 
and scalars (the transverse fluctuations) which transform as
\bea
             \mbox{Fermions}  & \ra&    8_{c}^{+1} \oplus 8_{s}^{-1}\non
             \mbox{Scalars}  &\ra & 8_{v}^{0}
\eea
under $Spin(8) \times U(1)$.  Another way of saying this is that
left-movers live in the $8_{c}$ and right-movers in the $8_{s}$. Modulo
the (locally trivial) gauge field, these are precisely the same
fluctuations as those of the fundamental IIB string, a manifestation of
the self-duality of the IIB string.

Clearly, two (partial) topological twists of the $N\!=\!8$ $d\!=\!2$ 
theory can be
obtained as dimensional reductions of the $N_{T}\!=\!2,4$ $d\!=\!3$ theories 
discussed in the previous section. As it turns out, these are precisely
the two theories we are after which describe the situations of interest, 
namely holomorphic curves in K3s and Calabi-Yau three-folds. 

Let us start with the $N_{T}\!=\!4$ $d\!=\!3$ theory. 
It follows from the description
of the theory given after (\ref{d3n4}) that the dimensionally reduced field
content is given by the gauge field, 
8 Grassmann odd scalars (hence this theory has
$N_{T}\!=\!8$), 4 Grassmann odd one-forms, six bosonic scalars and one 
bosonic one-form. Exhibiting their quantum numbers under the symmetry
group $SU(2)\times SU(2)$ manifest in (\ref{d3n4}) (we will see later that
this theory actually has a larger $SU(4)$ global symmetry), we see that
this topological twist amounts to
\bea
8_{c}^{+1} \oplus 8_{s}^{-1} &\ra& 2(2,1)^{0} \oplus 2 (1,2)^{0}
                  \oplus (2,1)^{+2,-2} \oplus (1,2)^{+2,-2}\non
8_{v}^{0} &\ra& (2,2)^{0} \oplus 2(1,1)^{0} \oplus (1,1)^{+2,-2}\;\;.
\label{d2n8}
\eea
Here the two $SU(2)\times SU(2)$ singlet scalars correspond to the
internal components $A_{3}$ and $V_{3}$ of the $d=3$ gauge field and
one-form.

In the same way, the 
$N_{T}\!=\!2$ theory (\ref{d3n2}) reduces to an $N_{T}\!=\!4$ theory with 
bosonic field content a $d\!=\!2$ gauge field, 4 scalars and two 
spinors, the twisting being described by 
\bea
8_{c}^{+1} \oplus 8_{s}^{-1} &\ra& 2 (1,2)^{0} \oplus (1,2)^{+2,-2}
  \oplus (2,2)^{+1,-1}\non
8_{v}^{0} &\ra& (2,1)^{+1,-1}\oplus (1,3)^{0} \oplus (1,1)^{0}\;\;.
\label{d2n4}
\eea

To find the geometrical interpretation or realization of these
theories, we need to know something about supersymmetric two-cycles.
As mentioned above, 
the only known supersymmetric 2-cycles are holomorphic curves in Calabi-Yau
$n$-folds. One might think that another candidate are special
Lagrangian submanifolds of a K3 (Calabi-Yau 2-fold). Indeed, at first sight
a special Lagrangian submanifold (in particular, the 
K\"ahler form restricts
to zero) and a holomorphic curve (the K\"ahler form restricts to the
volume form)
appear to be very different. However, K3's are actually hyper-K\"ahler
so that there exists a triplet $I,J,K$ of complex structures and it turns 
out \cite{wolfson} that a curve is special Lagrangian
with respect to  $I$ iff it is holomorphic with respect to $K$, so that 
indeed the special Lagrangian case need not be considered seperately. 

As before, identifications of the twisted world-sheet theories require
considerations of the normal bundle. A preliminary consideration is the
following: the manifolds we are dealing with are Calabi-Yau
$n$-folds $M_{n}.$ Hence
they satisfy $c_{1}(T_M)=0$ where $T_M$ is the holomorphic tangent bundle. 
Thus, for a holomorphic genus $g$ curve $S$ in $M_{n}$ one has
\bea
&&           T_{M}|_{S}=T_S \oplus N_S\non
&&           c_{1}(N_S) = -c_{1}(T_S) = 2g-2\;\;.
\eea
Alternatively, one notes that $\wedge^{n}T_M$ is trivial, so that
one has
\be
          1 = \wedge^{n}T_M|_{S} = T_S\wedge^{n-1}N_S
\ee
which gives the constraint 
\be
              \wedge^{n-1}N_S = K_S\;\;,\label{nsks}
\ee
$K_S$ the canoncial line bundle on $S$ (as $T_S=K_S^{-1}$).

Let us quickly dispose of the trivial case $n=1$.
Then $M$ is just a torus $T^{2},$ so that what we
are looking at is `compactification' on $T^{2}$ and a D-string whose
world-sheet is that $T^{2}$. The normal bundle is trivial, so the 
scalar field content should be eight scalars. In addition, no 
further supersymmetries are broken by this compactification, so 
the world-sheet theory is just Euclidean $N\!=\!8$ $d\!=\!2$ SYM with 
$N_{T}\!=\!16$
real supercharges (switching freely between supersymmetric and topological
notation as it makes no difference on $T^{2}$). 

When $n=2$, we are dealing with holomorphic curves in a K3. This case has
essentially been studied from this point of view in \cite{bsv}, cf.\ also
\cite{yz}.
$N_S$ has complex rank one and the condition (\ref{nsks}) is solved
uniquely by $N_S=K_S$. Compactification on K3 leaves
6 flat directions so we expect the scalar field content to be 6 
scalars and a one-form. This is just the field content of the $N_{T}\!=\!8$
model (the dimensional reduction of the fully topological  
$N_{T}\!=\!4$ $d\!=\!3$
theory). As K3 breaks half the supersymmetries, this matches with
$N_{T}=8$.

However, according to the general argument put forward in the previous
section, the six scalars should transform as a vector of $SO(6)$ which
should be a global symmetry group of this theory - whereas only 
$SU(2)\times SU(2)$ is manifest in (\ref{d2n8}). To exhibit this
$SO(6)\sim SU(4)$ symmetry of the theory, we proceed as follows. 
We consider the branching (cf.\ (\ref{geob}))
\be
Spin(8) \ra SU(4) \times U(1)\;\;,
\ee
under which the representations $8_{v,s,c}$ decompose as
\bea
8_{v} &\ra& (1)^{+2,-2}\oplus (6)^{0}\non
8_{s} &\ra& (4)^{+1} \oplus (\bar{4})^{-1}\non
8_{c} &\ra& (4)^{-1} \oplus (\bar{4})^{+1}\;\;.
\eea
Thus, twisting the Lorentz $U(1)$-charge by the internal $U(1)$-charge
(by simply adding them up) and leaving the $SU(4)$ intact, one finds
that the field content of the twisted theory is
\bea
8_{c}^{+1}\oplus 8_{s}^{-1} &\ra& 2(4)^{0} \oplus (\bar{4})^{+2,-2}\non
8_{v}^{0} &\ra& (6)^{0} \oplus (1)^{+2,-2}\;\;.
\eea
This has the expected manifest global $SU(4)$ symmetry under which the 
six scalars transform as a vector, and it reduces to the field content 
with the quantum numbers as in (\ref{d2n8}) under the further branching 
\be
SU(4) \ra SU(2)\times SU(2)\times U(1)\;\;.
\ee

When $n=3$, we are dealing with holomorphic curves in Calabi-Yau 3-folds. 
This has been thoroughly studied (for a
recent review see \cite{morrison}), and our brief presentation of the
results will certainly not do justice to the complexity and 
diversity of the subject. 

In this case, the condition (\ref{nsks}) on $N_S$ reads $\wedge^{2}N_S=K_S$  
and generally this is solved by
\be
         N_S = K_{S}^{1/2} \otimes V,\label{nsc3}
\ee
where $V$ is a rank two bundle with trivial determinant. 
This includes as a special case unobstructed rational curves for which
one has 
\be
T_{M}|_{S}={\cal O}_{S}(2) \oplus {\cal O}_{S}(-1)\oplus {\cal
O}_{S}(-1)\;\;,
\ee
and for which $V$ is trivial, as well as other situations such as
$V=K_S^{1/2}\oplus K_S^{-1/2}$, which could appear for $M_{3}=K3 \times T^{2}$
(we will briefly come back to that case below). Thus 
the bosonic field content of the world-sheet theory of the D-string instanton
wrapping around such a holomorphic curve should consist of 4 scalars (for
the uncompactified directions) and a doublet of spinors. This is exactly the
field content of the $N_{T}\!=\!4$ theory (\ref{d2n4}), the dimensional 
reduction of the
half-twisted $N_{T}\!=\!2$ $d\!=\!3$ theory, and again the supersymmetries work
out as a Calabi-Yau 3-fold breaks 1/4 of the supersymmetries, leaving one 
with $N_{T}=16/4=4$. Geometrical considerations, as in (\ref{geob}), 
suggest that this twisted theory should arise from the branching
$SO(8)\ra SO(4)\times SO(4)$ or
\be
Spin(8)\ra SU(2)\times SU(2)\times SU(2)\times SU(2)\;\;,
\ee
and it can indeed be seen that the corresponding branchings 
\bea
8_{v}&\ra& (2,2,1,1)\oplus (1,1,2,2)\non
8_{s}&\ra& (1,2,1,2)\oplus (2,1,2,1)\non
8_{c}&\ra& (1,2,2,1)\oplus (2,1,1,2)
\eea
reproduce the field content (\ref{d2n4}) with an extended $SO(4)\times SU(2)$
global symmetry corresponding to rotations in the uncompactified dimensions
and on the internal $SU(2)$ associated with the vector bundle $V$.

Let us now come back to the special case where the Calabi-Yau
three-fold is of the form $K3\times T^{2}$ with holonomy $SU(2)\ss SU(3)$.
In that case, one would expect an enlarged topological symmetry to be
present. This is indeed the case and comes about as follows.

If e.g.\ $S\ss K3$, then its normal bundle in $K3$ can be identified with
$K_{S}$ and thus its normal bundle in $K3\times T^{2}$ is 
\be
N_{S} =K_{S}\oplus {\cal O}_{S}\;\;.
\ee
This is of the form
(\ref{nsc3}) for $V=K_S^{1/2}\oplus K_{S}^{-1/2}$. Thus the structure
group of $V$ has been reduced to $U(1)$. The identification 
$K_{S}^{1/2}\otimes V\sim K_{S}\oplus {\cal O}_{S}$ is then realized
on the fields by a further twisting
by this $U(1)$. As usual, this turns the $SU(2)$ spinor doublet into
a one-form and two scalars. Its effect on the Grassmann odd field
content displayed in the first line of (\ref{d2n4}) is to turn it into 
eight Grassmann odd scalars and four one-forms. Thus this twisted model
shows the expected increase in the number of topological symmetries
from $N_{T}\!=\!4$ to $N_{T}\!=\!8$.  
In fact, not too surprisingly, this model coincides with the 
$N_{T}=8$ model (\ref{d2n8}), the difference between $\RR^{2}$ and
$T^{2}$ (i.e.\ the fact that the deformations are compact for the latter)
being invisible in the low-energy effective action. Similar remarks
apply to other situations in which the holonomy group of the $n$-fold
is a strict subgroup of $SU(n)$. 

\subsubsection*{Acknowledgements}

G.T.\ would like to thank the University of Paris 7 for a visiting
Professorship during which period this work was begun. He would also like 
to thank the members of the LPTHE for their kind hospitality during his stay.

\rnc{\Large}{\normalsize}

\end{document}